\documentclass[preprint,5p,sort&compress]{elsarticle}
\usepackage{color}
\usepackage{multirow}
\usepackage{mathptm}
\usepackage{amsfonts}
\usepackage{amsmath}
\usepackage{amssymb}
\usepackage{graphicx}
\usepackage{sectsty}
\usepackage{bbold}
\usepackage{subfigure}
\usepackage{algorithm}
\usepackage{algorithmicx}
\usepackage{algpseudocode}

\newcommand\be{\begin{equation}}
\newcommand\ee{\end{equation}}
\newcommand\bea{\begin{eqnarray}}
\newcommand\eea{\end{eqnarray}}
\newcommand\bi{\begin{itemize}}
\newcommand\ei{\end{itemize}}
\newcommand\nn{\nonumber}
\newcommand\Ham{\hat{H}}

\newcommand\nr{{\it n}({\bf r})}
\newcommand\kv{{\bf k}}
\newcommand\rv{{\bf r}}
\newcommand\blk{{\tt blk}}
\newcommand\polym{{\tt deg}}
\newcommand\nev{{\tt nev}}

\newcommand{\seq}{$\{P^{(\ell)}\}$}

\newcommand\eqalign[1]{%
	\vcenter{%
		\normalbaselines \advance\baselineskip 5pt
		\advance\lineskip 5pt \tabskip=0pt
		\halign{%
			&\hfil $\displaystyle{##{}}$&
			$\displaystyle{{}##}$\hfil\cr
			#1\crcr
			}%
		}%
	}

\newcommand\dotline{\par\hbox to \hsize{\dotfill}\par}

\makeatletter
\def\befored@t#1.#2.#3;{#1}
\def\afterd@t#1.#2.#3;{#2}
\def\refhead#1{\edef\next{\ref{#1}}\expandafter\befored@t\next..;}
\def\reftail#1{\edef\next{\ref{#1}}\expandafter\afterd@t\next..;}
\def\lsim{\mathrel{\mathpalette\@versim<}}
\def\gsim{\mathrel{\mathpalette\@versim>}}
\def\@versim#1#2{\vcenter{\offinterlineskip
        \ialign{$\m@th#1\hfil##\hfil$\crcr#2\crcr\sim\crcr } }}
\newcommand\becomes[1]{\mathchoice{\becomes@\scriptstyle{#1}}
   {\becomes@\scriptstyle{#1}} {\becomes@\scriptscriptstyle{#1}}
   {\becomes@\scriptscriptstyle{#1}}}
\def\becomes@#1#2{\mathrel{\setbox0=\hbox{$\m@th #1{\,#2\,}$}%
        \mathop{\hbox to \wd0 {\rightarrowfill}}\limits_{#2}}}
\makeatother
\begin{document}
%
%
\title{Block Iterative Eigensolvers for Sequences of Correlated Eigenvalue Problems\tnoteref{t1,t2}}
\tnotetext[t1]{Article based on research supported by the VolkswagenStiftung through the``Computational Science'' fellowship}
\tnotetext[t2]{Preprints: AICES--2012/12--1 \quad ArXiv:1206.3768v1}

\author[ads1,ads2]{Edoardo Di Napoli\corref{cor1}}
\ead{e.di.napoli@fz-juelich.de}

\author[ads3]{Mario Berljafa}
\ead{mberljaf@student.math.hr}

\address[ads1]{J\"ulich Supercomputing Centre, Institute for Advanced Simulation, Forschungszentrum J\"ulich, Wilhelm-Johnen strasse, 52425 J\"ulich, Germany}
\address[ads2]{RWTH Aachen University, AICES, Schinkelstr. 2, 52062 Aachen, Germany}
\address[ads3]{Department of Mathematics, Faculty of Science, University of Zagreb, Bijeni\v{c}ka cesta 30, 10000 Zagreb, Republic of Croatia.}

\cortext[cor1]{Principal corresponding author}


\begin{abstract} 
  In Density Functional Theory simulations based on the LAPW method,
  each self-consistent field cycle comprises dozens of large dense
  generalized eigenproblems. In contrast to real-space methods,
  eigenpairs solving for problems at distinct cycles have either been
  believed to be independent or at most very loosely connected.  In a
  recent study~\cite{DBB}, it was demonstrated that, contrary to
  belief, successive eigenproblems in a sequence are strongly
  correlated with one another. In particular, by monitoring the
  subspace angles between eigenvectors of successive eigenproblems, it
  was shown that these angles decrease noticeably after the first few
  iterations and become close to collinear. This last result suggests
  that we can manipulate the eigenvectors, solving for a specific
  eigenproblem in a sequence, as an approximate solution for the
  following eigenproblem.  In this work we present results that are in
  line with this intuition. We provide numerical examples where
  opportunely selected block iterative eigensolvers benefit from the
  reuse of eigenvectors by achieving a substantial speed-up. The
  results presented will eventually open the way to a widespread use
  of block iterative eigensolvers in ab initio electronic structure
  codes based on the LAPW approach.
\end{abstract}


\maketitle
 
\pagenumbering{arabic}
%
%
\section{Introduction}
\label{sec:intro}

Materials simulations based on Density Functional Theory~\cite{DG} (DFT)
methods have at their core a set of partial differential equations (Kohn-Sham
\cite{KS}) which eventually lead to a non-linear generalized
eigenvalue problem. Solving the latter directly is a daunting task
and a numerical iterative self-consistent approach is preferred.
It starts off by inputting an approximate electronic charge density 
to a cyclic process within which a linearized version of the eigenvalue problem
is initialized and solved. At the end of each cycle a new charge density is computed 
and compared with the initial one. Self-consistency is reached when the distance
between the input and output densities is below a certain required
threshold; the process is then said to have converged.
The entire simulation results in a series of so called ``outer-iteration cycles'' often referred
to as self-consistent field iterations. 

Roughly speaking, all the existing DFT-based methods differ from each
other by the choice of linearization scheme (also denoted as
discretization), and by the choice of the effective Kohn-Sham (KS)
potential.  There are three discretization strategies commonly in use:
1) manipulation of localized functions (Gaussians, etc.), 2) expansion
of the eigenfunctions over a plane wave basis set, and 3)
discretization of the KS equations over a lattice in real space.
While the first method is almost exclusively used in Quantum Chemistry
the last two are widely used in Materials Science and present a series
of pros and cons. The plane wave expansion leads to Hamiltonians with
kinetic energy terms only on the main diagonal and are well suited to
simulate solid crystals. In turn this discretization needs to
approximate the Coulomb potential near the nuclei substituting it with
a smooth pseudo-potential. In real-space discretization, potential
terms in the Hamiltonian decay exponentially away from the
diagonal~\cite{Benzi} giving rise to quite sparse and large eigenvalue
problems. This strategy is well suited mostly for disordered systems
and insulators.
 
Among the plane wave strategies, the Full-Potential Linearized
Augmented Plane Wave (FLAPW) ~\cite{FKWW,FJ} method constitutes one of
the most precise frameworks for simulating transition metals and
magnetic systems. The Kohn-Sham equations are discretized using a mix
of radial and plane wave functions (see section~\ref{sec:SeqDens}),
parametrized by a vector \kv\ within the Brillouin zone of the
momentum lattice.  At each outer-iteration $\ell$ a set of
eigenpencils $P^{(\ell)}_\kv$, labeled by \kv , is initialized and
solved. Because FLAPW uses full-potential together with a partial
plane wave expansion, each $P^{(\ell)}_\kv$ is a dense and hermitian
generalized eigenvalue problem; its size $n$ depends linearly -- with
a large pre-factor -- on the number of atoms considered in a
simulation, and typically ranges between 2,000 and 20,000. Only a
relatively small percentage of the bottom end of the spectrum is
required, never exceeding 15-20\%, and often quite less.

In this work we consider sequences of generalized hermitian eigenvalue problems
as they arise in FLAPW. In this context a sequence is 
a set of $N$ generalized eigenproblems  identified by a progressive index $\ell$ 
\be
\label{eq:seq}
	\{P^{(\ell)}\} \doteq P^{(1)},\dots, P^{(N)} \qquad ; \qquad P^{(\ell)}:\quad A^{(\ell)} x = \lambda\ B^{(\ell)} x .
\ee
Within a sequence
each eigenproblem is characterized by a hermitian indefinite matrix $A$
and a positive definite hermitian matrix $B$. This setup is generally referred to as
a matrix pencil or eigenpencil and it is known to have a bounded discrete spectrum with real positive and negative
eigenvalues
\be
	\lambda_{\rm min} = \lambda_1 \leq \lambda_2 \leq \dots \leq \lambda_n = \lambda_{\rm max}.
\ee 
Eigenpencils usually admit $n$ distinct eigenvectors $x_i$ satisfying a $B$-orthonormality relation
$(x_i, Bx_j) = \delta_{ij}$ even when they correspond to identical eigenvalues. 
While in general $B \neq I$, in the special case $B = I$ the eigenpencil becomes a standard 
eigenvalue problem, and the orthonormality relation reduces to the standard $(x_i, x_j) = \delta_{ij}$.

In current codes implementing
FLAPW~\cite{FLEUR,Wien2k,FLAIR,Exciting,Elk}, each sequence of
eigenpencils \seq\ is handled very much as a set of uncorrelated
problems: each $P^{(\ell)}$ is solved in complete isolation from any
other and independently passed as input to a prepackaged eigensolver
of a standard library -- like LAPACK~\cite{LAPACK3} or its parallel
version ScaLAPACK~\cite{ScaLAPACK} -- which outputs the desired
portion of eigenspectrum and corresponding eigenvectors.  The
eigensolver is thus used as a black box and has no knowledge of the
eigenproblems' spectral properties nor of the application from which
they originated.  As much as this process grants standardization and
reliability, it is also far from being optimal. What is ``lost in
translation'' is the possibility to render manifest the correlation
between eigenpencils of the sequence \seq\ in terms of precise
numerical properties which are then passed to a solver that can
exploit them.

In a recent work~\cite{DBB} it has been reported that eigenpencils
with a successive outer-iteration index $\ell$ and the same \kv-vector
are strongly correlated.  Consequently, problems in a sequence are not
only connected by a progressive index but, as for a sequence of
numbers, there is a relation linking them.  In FLAPW, such a numerical
correlation become evident in the way the subspace angles between
eigenvectors evolve from larger to smaller values as the sequence
progresses towards higher outer-iteration indices~\cite{DBB}. It needs
to be stressed that, contrary to what happens in real-space methods,
the correlation between eigenvectors is a new and unexpected feature
of FLAPW-based methods: since the eigenfunctions are delocalized and
the function basis set is modified at each successive outer-iteration,
it had been common belief that correlation was an unlikely phenomenon.

With evidence of the contrary in hand, it becomes natural to consider
eigenvectors of $P^{(\ell)}$ as a set of approximate solutions that
can be used by an appropriate eigensolver to accelerate the solution
of $P^{(\ell+1)}$.  The novelty of our contribution consists in
showing that, by exploiting the collinearity between vectors of
adjacent problems, we can significantly improve the performance of
certain classes of eigensolvers. Since no eigensolver (QR, MRRR,
Divide\&Conquer, etc.) for dense problems accepts as input approximate
eigenvectors, our strategy can only be carried out by using iterative
eigensolvers. In the rest of the paper we first illustrate, through
numerical experiments, the success of this strategy for three distinct
block iterative eigensolvers, each representing a specific class of
available methods. Then we focus on one of these solvers, develop a C
language version and obtain similar results with emphasis on
high-performance and scalability.

In section~\ref{sec:SeqDens}
we first give a short description of how sequences of eigenproblems arise
in DFT and how they translate into apparently uncorrelated dense eigenvalue problems.
We then proceed to briefly report on the correlation between adjacent eigenproblems
as illustrated in~\cite{DBB}.
Our core results are presented in section~\ref{sec:evol} where we introduce
the selected block iterative eigensolvers followed by a description of the experimental setup 
and the numerical tests performed. We summarize our results in section~\ref{sec:sum},
and conclude with future work and acknowledgments.


  \section{Sequences of correlated eigenproblems}
  \label{sec:SeqDens}
  
  In this section we illustrate in some detail how sequences of eigenpencils
  arise in DFT. We start with a brief recall of the fundamentals of quantum mechanics,
  explain the need for an effective theory dealing with many particles
  and describe the FLAPW method self-consistent cycle. It is then shown why
  correlation among eigenproblems in a sequence is unexpected, and how
  the presence of such a correlation was exposed by looking at the evolution of eigenvectors
  as a function of the outer-iteration cycle index $\ell$. 
  
  
  \subsection{The rise of sequences in Density Functional Theory}
  \label{sec:rise}     
  The electronic structure of a quantum mechanical system with $L$ atoms and $M$ electrons 
  is described by the Schr\"odinger equation
  \be
  \label{eq:schr}
  	H\, \Phi(x_1;s_1,\dots ,x_n;s_n) = \mathcal{E}\, \Phi(x_1;s_1,\dots ,x_n;s_n).
  \ee
  $H= - \frac{\hbar^2}{2m} \sum_{i=1}^M \nabla^2_i - \sum_{i=1}^M \sum_{\mu=1}^L 
  \frac{Z_\mu}{|x_i - a_\mu|} + \sum_{i < j} \frac{1}{|x_i - x_j|}$ is the Hamiltonian characterizing the 
  dynamics of the electrons whose positions and spins are indicated by $x$ and 
  $s$ respectively. $\mathcal{E}$ represents the energy of the system while $\Phi$ is the high-dimensional 
  antisymmetric electronic wave function solving for eq.~(\ref{eq:schr}). Already at this stage the Schr\"odinger
  equation looks very much like an eigenvalue problem, unfortunately one that is already very 
  challenging to solve for values $M,L \geq 2$.
    
  During the 1960s, a series of simplifications were introduced based on rigorous theorems~\cite{KS, HK}
  where the exact high-dimensional eq.~(\ref{eq:schr}) was replaced by a large set of one-dimensional Kohn-Sham equations
  \be
  \label{eq:KS}
	\forall\ a \quad \textrm{solve} \quad \Ham_{\rm KS} \phi_a({\bf r}) = \left( -\frac{\hbar^2}{2m} \nabla^2 + V_0({\bf r}) \right) \phi_a({\bf r}) = \epsilon_a \phi_a({\bf r}). 
  \ee
  The most important element in these equations is the substitution of the last two terms of $H$ with an effective
  potential $V_0({\bf r})[n]$ that functionally depends on the charge density $n({\bf r})$: a function of all the one-particle
  wave functions $\phi_a({\bf r})$. Because of this interdependence between $V_0$ and $\phi_a({\bf r})$, eq.~(\ref{eq:KS})
  constitutes a set of non-linear partial differential equations.
    
  Typically this set of equations is solved using an outer-iterative
  self-consistent cycle: it starts off with an initial charge density
  $n_{init}({\bf r})$, proceeds through a series of iterations and
  converges to a final density $n_N({\bf r})$ such that $|n^{(N)} -
  n^{(N-1)}| < \eta$, with $\eta$ as an a priori
  parameter. Convergence is achieved by an oppurtune mixing between
  output density $n_i({\bf r})$ and one or more previous input
  densities $n_{\ell < i}({\bf r})$. In the particular case of FLAPW
  the new density is computed by a quasi-Newton mixing with the
  inverse Jacobian updated by Broyden's second method~\cite{johnson,blugo}.
  \begin{equation}
  \label{eq:cycle}
	\begin{array}[l]{ccc}
 	\color{red}{\fbox{\rule[-0.2cm]{0cm}{0.6cm}\shortstack{{\small Initial input}\\$n_{\rm init}(\rv)$}}} &  &  \\[4mm]
	 \Downarrow &  & \\[2mm]
	\fbox{\rule[-0.2cm]{0cm}{0.6cm} \shortstack{{\small Compute KS potential}\\$V_0(\rv)[n]$}} & 
	\longrightarrow & \fbox{\rule[-0.2cm]{0cm}{0.6cm} \shortstack{{\small Solve KS equations}\\$\Ham_{\rm KS}\phi_a({\bf r}) = \epsilon_a \phi_a({\bf r})$}} \\[4mm]
 	 \uparrow {\rm No} &  & \downarrow  \\[2mm]
	 \fbox{\rule[-0.2cm]{0cm}{0.6cm} \shortstack{\small Converged? \\ $|n^{(\ell+1)} - n^{(\ell)}| < \eta$}} & 
	 \longleftarrow &  \fbox{\rule[-0.2cm]{0cm}{0.6cm} \shortstack{{\small Compute new density}\\$n(\rv) = \sum_a |\phi_a({\bf r})|^2$}}\\[4mm]
	 \Downarrow \textrm{Yes} &  & \\[4mm]
	 \color{blue}{\fbox{\rule[-0.2cm]{0cm}{0.6cm} \shortstack{{\small OUTPUT}\\ {\small Energy, forces, etc.}}}} &  &
 	\end{array}
  \end{equation}

  In practice this outer-iterative cycle is still quite computationally challenging and requires some form of broadly defined discretization.
  In the FLAPW method, the wave functions $\phi_a({\bf r})$ are expanded on a basis set $\psi_{\bf G}(\kv,\rv)$ parametrized by vectors \kv\ living
  in the momentum space discretized on a lattice
  \be
  \label{eq:expon}
	\phi_a({\bf r}) \longrightarrow \phi_{\kv,\nu}(\rv) = \sum_{|{\bf G + k}|\leq {\bf K}_{max}} c^{\bf G}_{\kv,\nu} \psi_{\bf G}(\kv,\rv). 
  \ee
  Here ${\bf K}_{max}$ is a cut-off and its value determines the range of the vector index ${\bf G}$, ultimately 
  controlling the size of the eigenproblems $n$ (not to be confused with the charge density \nr). Thus, the number of basis functions in the expansion equals the size of the
  problem at hand. In FLAPW the basis functions $\psi_{\bf G}(\kv,\rv)$ are constructed by merging together radial-like 
  functions $u^{\alpha}_{\it l}(r)$ inside
  a spherical region around the atoms (also called Muffin-Tin) and simple plane waves in the interstitial areas between atoms
  \begin{eqnarray*}
  \lefteqn{\psi_{\bf G}(\kv,\rv) = }&  \\
  \hspace{-0.5cm} = & \hspace{-0.3cm} \left\{
	\begin{array}[l]{l}
	\frac{1}{\sqrt{\Omega}}e^{i({\bf k+G})\rv} \qquad \qquad \qquad \qquad \qquad \quad -\textrm{Interstitial}\\
	\displaystyle\sum_{\it l,m} \left[a^{\alpha,{\bf G}}_{\it lm}(\kv) u^{\alpha}_{\it l}(r) 
	+ b^{\alpha,{\bf G}}_{\it lm}(\kv) \dot{u}^{\alpha}_{\it l}(r) \right] Y_{\it lm}(\hat{\bf r}_{\alpha}) \hfill - \textrm{MT}. \\
	\end{array}
  \right.  \\
  \end{eqnarray*}

At each iteration cycle the radial functions are computed anew by solving auxiliary Schr\"odinger equations.
Moreover a new set of the coefficients $a^{\alpha,{\bf G}}_{\it lm}$ and 
$b^{\alpha,{\bf G}}_{\it lm}$ is derived by imposing continuity constraints on the surface of the Muffin-Tin spheres.
Consequently at each iteration the basis set $\psi_{\bf G}(\kv,\rv)$ changes entirely. As will be explained in the next section, this peculiarity
of the FLAPW method is one of the main reasons why correlation between eigenvectors of adjacent eigenproblems
had been, until recently, considered unlikely. 
  
Having defined a basis set of wave functions allows us to translate the KS equations into a set of generalized eigenvalue problems. First the expression (\ref{eq:expon}) is inserted in eq.~(\ref{eq:KS})
  \be
  \label{eq:eigenstep}
		\psi^*_{\bf G}(\kv,\rv) \sum_{\bf G'} \Ham_{\rm KS}\ c^{\bf G'}_{\kv,\nu}\ \psi_{\bf G'}(\kv,\rv) = \lambda_{\kv\nu}\ \psi^*_{\bf G}(\kv,\rv) \sum_{\bf G'} c^{\bf G'}_{\kv,\nu}\ \psi_{\bf G'}(\kv,\rv). 
  \ee
  Then, by integrating the left and right hand side of eq.~(\ref{eq:eigenstep}) over the configuration space, the matrix entries for the Hamiltonian $A_\kv$ and Overlap matrices $B_\kv$ are computed
  \be
  \label{eq:matrices}
		\left[A_\kv \,\, B_\kv\right]_{\bf G G'} = \sum_\alpha \int d{\bf r}\  \psi^{\ast}_{\bf G}(\kv,\rv) \left[\Ham_{\bf KS}\,\, \hat{\mathbb{1}}\right]  \psi_{\bf G'}(\kv,\rv).
  \ee
  The end result is a set of generalized eigenvalue equations parametrized by the vector \kv 
  \be
	P_\kv: \quad \sum_{\bf G'} (A_\kv)_{\bf GG'}\ c^{\bf G'}_{\kv\nu} = \lambda_{\kv\nu} \sum_{\bf G'} (B_\kv)_{\bf GG'}\ c^{\bf G'}_{\kv\nu}
	\hfill \,\, \equiv \,\, A_\kv x_i = \lambda_i B_\kv x_i. \nn
  \ee
  As can be readily seen the coefficients $c_{\kv\nu}$ play the role of eigenvectors while the indices $\kv$ and $\nu$ can be compactly condensed 
  in the single index $i$. Moreover because of the complex structure of $\psi_{\bf G}(\kv,\rv)$, interactions in $\Ham_{\bf KS}$ are ``delocalized''
  with the net effect of filling up the Hamiltonian matrix $A$ and losing any diagonal dominance. At the same time the over-completeness 
  of the basis set renders $B$ dense as well as somewhat ill-conditioned. These characteristics differentiate FLAPW-based methods from real-space ones,
  and make them very well suited to simulate physical systems where electrons are partially delocalized (metals, semi-conductors, etc.).
   
  The net effect of this discretization is the translation of the KS equations into a set of generalized eigenvalue problems for each outer-iteration
  \be
	\begin{array}[l]{ccc}
 	\fbox{\rule[-0.2cm]{0cm}{0.6cm} \shortstack{{\small Solve KS equations}\\$\Ham_{\rm KS}\phi_a({\bf r}) = \epsilon_a \phi_a({\bf r})$}} &
	\Rightarrow & \fbox{\rule[-0.2cm]{0cm}{0.6cm} \shortstack{{ Solve a set of eigenproblems}\\$P_{\kv_1} \dots P_{\kv_N}$}}.
         \end{array}
         \nn
  \ee
  In the end the entire process has at its core the initialization and solution of a set of sequences of dense eigenpencils $\{P^{(\ell)}_\kv\}$. 


  \subsection{Correlation in the sequence}
  \label{sec:corr}
  In analogy to the definition of a sequence of numbers,
  a sequence of eigenvalue problems is considered
  as such if $P^{(\ell)}$ has some connection to $P^{(\ell+1)}\ \forall \ell$. 
  Therefore, we now focus on a single sequence \seq\ and describe how 
  this correlation becomes manifest in the evolution of subspace deviation angles
  between eigenvectors of successive eigenproblems. 
  
  In sequences arising from DFT the most obvious form of connection comes
  from the fact that the problem with index $\ell+1$ is initialized only after the previous
  problem is solved. Despite this simple fact, the initialization and solution of the new eigenproblem
  depends very indirectly on the solution of the old one. Eq.~(\ref{eq:matrices}) shows that
  matrix entries are computed using a new set of $\psi^{(\ell+1)}_{\bf G}(\kv,\rv)$, whose Muffin-Tin (MT)
  components can be radically different from the one of $\psi^{(\ell)}_{\bf G}(\kv,\rv)$.
  The same MT components depend very non-linearly on the charge density
  $\nr$ which, in turn, is a sum of linear combinations (see eq.~(\ref{eq:expon})) of eigenvectors
  $c^{(\ell)}_{\kv\nu}$ and the same basis functions $\psi^{(\ell)}_{\bf G}(\kv,\rv)$.
  In conclusion at each outer-iteration, the Hamiltonian, the Overlap and the basis function set
  vary substantially and in a way that cannot be estimated. 
  Due to this series of deductions, the correlation between successive
  eigenvalue problems in FLAPW is a rather novel discovery which had been considered
  by the computational physics community to be a rather unlikely outcome.
  
  In \cite{DBB} the link between adjacent eigenproblems is unraveled by looking
  at the existence of a correlation between eigenvectors as a numerical inverse problem.
  Starting from a constrained assumption, the authors conduct an a posteriori
  numerical analysis of angles between eigenvectors with successive outer-iteration indices $\ell$.
  For instance, let us look at two neighboring eigenproblems of a sequence 
  \be
  A^{(\ell)} x^{(\ell)} = \lambda\ B^{(\ell)} x^{(\ell)} \quad {\rm and} \quad
  A^{(\ell+1)} x^{(\ell+1)} = \lambda\ B^{(\ell+1)} x^{(\ell+1)} \nn,
  \ee 
  with $B^{(\ell)} = L^{(\ell)}\ L^{(\ell)T}$ and
  $B^{(\ell+1)} = L^{(\ell+1)}\ L^{(\ell+1)T}$ the Cholesky decomposition of the
  respective Overlap matrices. In~\cite{DBB} it is shown that the eigenvectors of these two problems
  satisfy the following general relation
  \begin{align}
  \label{eq:exp}
		\langle x^{(\ell)}_i x^{(\ell+1)}_j \rangle =  &\ c^{*(\ell)}_{\kv,\nu} L^{(\ell)}\ L^{(\ell+1)T}\ c^{(\ell+1)}_{\kv,\mu} \\
		=&\ \delta_{ij} + \varepsilon \left[ E_{ij} - c^{*(\ell)}_{\kv,\nu}\ L^{(\ell)}\ D\ L^{(\ell+1)T}\ c^{(\ell+1)}_{\kv,\mu} \right] + o(\varepsilon^2), \nn
  \end{align}
  where both $E$ and $D$ are block diagonal, each block being diagonally dominant, and
  $\varepsilon$ being an expansion parameter.
  As long as $\varepsilon$ is small, eq.~(\ref{eq:exp}) implies that the matrix of scalar products between eigenvectors of
  adjacent eigenproblems has a lumpy structure with most of its dominant
  entries concentrated in the neighborhood of the main diagonal. This structure
  makes it possible to devise an algorithm establishing a one-to-one
  association between a generic  eigenvector
  $x^{(\ell)}_i$ of $A^{(\ell)}$ and the corresponding eigenvector 
  $x^{(\ell+1)}_i$ of $A^{(\ell+1)}$. 
  
\begin{figure}[!htb] 
  \centering
	\subfigure[Lowest 3\% of eigenspectrum for CaFe$_2$As$_2$.]
	{\includegraphics[scale=0.23]{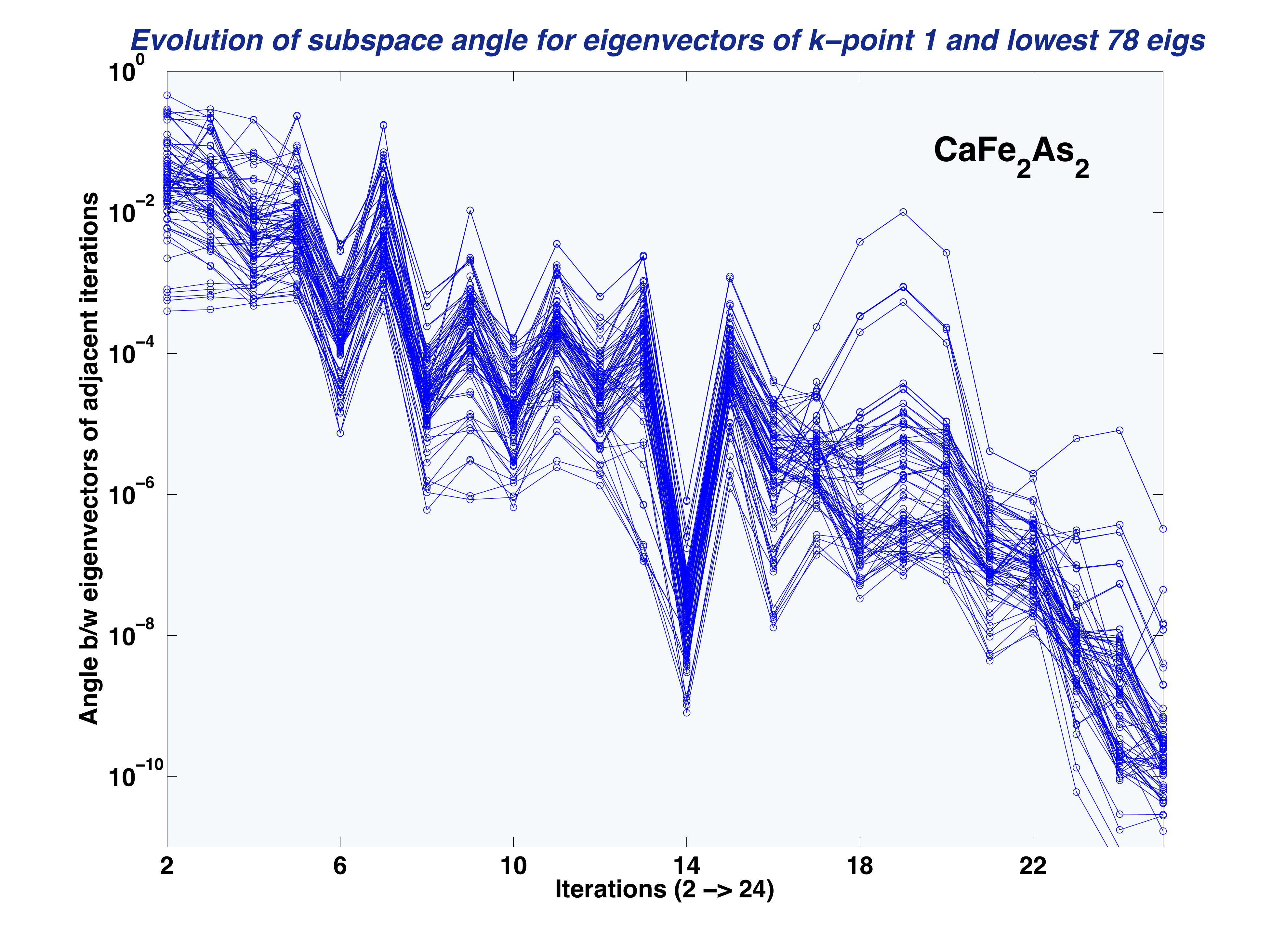}}
	\subfigure[Lowest 1.5\% of eigenspectrum for Au$_{98}$Ag$_{10}$.]
	{\includegraphics[scale=0.23]{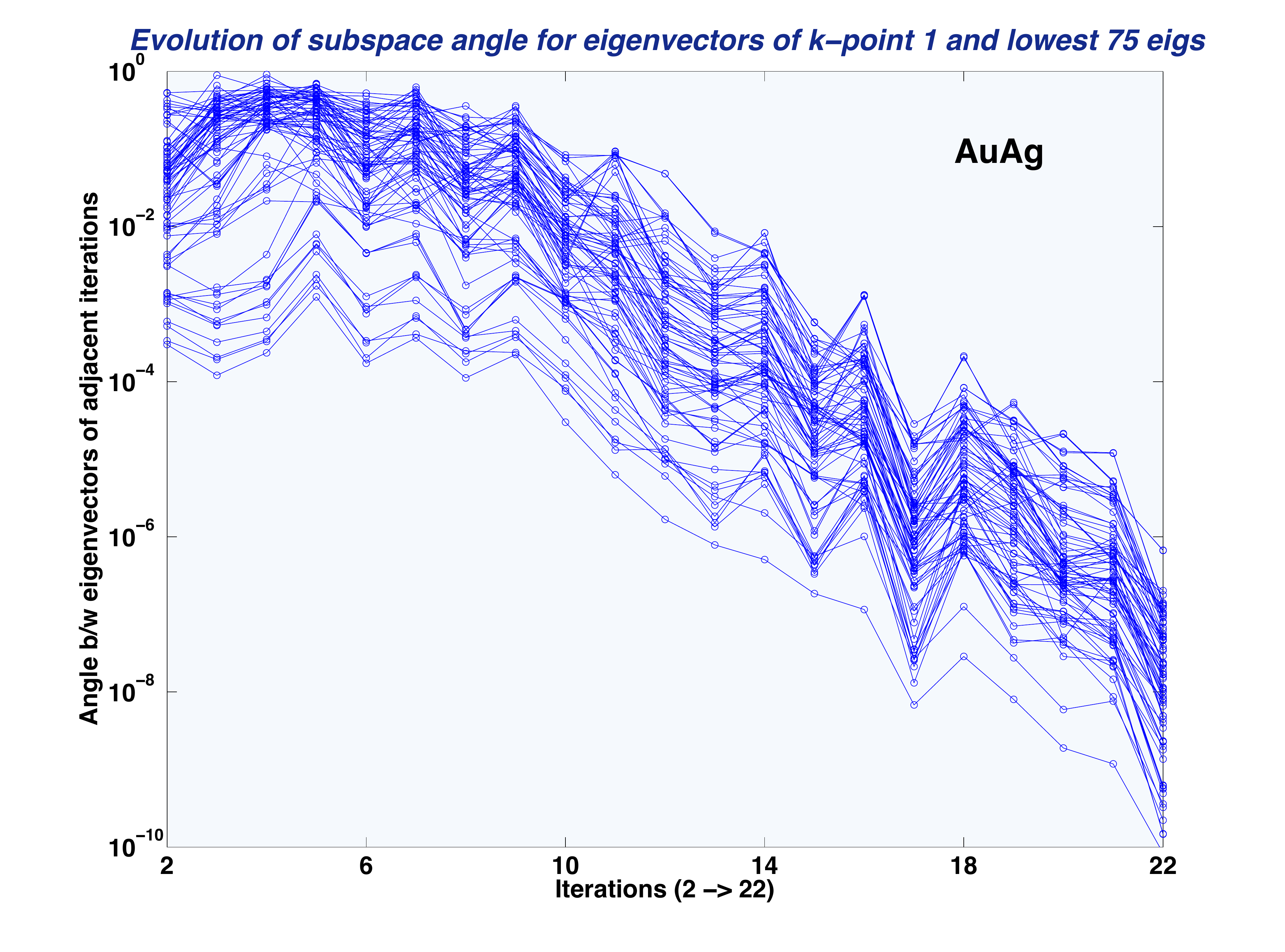}}
	\caption{\em Evolution of subspace deviation angles between eigenvectors of adjacent eigenvalue problems within a sequence.}
	\label{fig:evol}
  \end{figure}

  Once established, the one-to-one correspondence lays the ground for
  a systematic numerical analysis of the distribution of deviation angles
  $\theta^{(\ell)}_i = \langle x^{(\ell)}_i x^{(\ell+1)}_i \rangle$
  along the entire sequence. Plots of $\theta^{(\ell)}_i$
  illustrate how the eigenvectors become more and more collinear 
  as the sequence progresses (e.g.~Figure~\ref{fig:evol}).
  In particular, we can observe how angles are 
  already very small -- $\sim$ 10$^{-4}$ on average -- 
  after a few iterations, becoming almost negligible towards the end of 
  the self-consistent cycle. This result confirms a posteriori the
  correctness and reliability of the expansion~(\ref{eq:exp}) in powers of $\varepsilon$.
  
  The importance of the eigenvector collinearity is two-fold:
  on the one hand it makes it clear that there is a deep connection between
  the convergence of the charge density $n({\bf r})$ and the solutions of
  the eigenpencils. At the same time it provides the means to go beyond
  the current algorithmic paradigm so as to improve the solving process
  of the entire sequence.
  
  \begin{table}[ht]
  \caption{ Simulation data}
  \vspace{0.2cm}
  \centering
  \begin{tabular}{c c c c c }
  \hline \hline \\[-3mm]%
  Materials & \nev\ & {\bf K}$_{\rm max}$ & \shortstack{\# of \\iterations} & \shortstack{Size of \\eigenproblems}\\ [0.5ex]
  \hline \\[-3mm]%
  CaFe$_2$As$_2$ & 136 & 4.0 & 24 & 2,612\\ \hline%
  \multirow{2}{*}{Au$_{98}$Ag$_{10}$} & \multirow{2}{*}{972} & 3.0 & 22 & 5,638\\ 
   & & 3.5 & 29 & 8,970\\ \hline%
  \multirow{3}{*}{Na$_{15}$Cl$_{14}$Li} & \multirow{3}{*}{256} & 3.0 & 13 & 3,893\\ 
  	& & 3.5 & 13 & 6,217\\
	& & 4.0 & 13 & 9,273\\
  \end{tabular}\\
  \label{tab:sim}
  \end{table}
  
  We will see in the next section how this is possible for a selected group of 
  iterative eigensolvers. We conclude this section by illustrating two examples
  of evolution of deviation angles taken from real case solid-state crystals: CaFe$_2$As$_2$
  and Au$_{98}$Ag$_{10}$. The first is a superconducting compound that undergoes 
  a first order phase transition from a high temperature tetragonal phase to a 
  low temperature orthorhombic phase. The latter is an artificial alloy of gold and
  silver. In order to avoid cluttering,
  in Figure~\ref{fig:evol} $\theta^{(\ell)}_i$ are plotted only for a small fraction 
  of the eigenspectrum. In both plots, we can observe the deviation angles 
  decreasing quasi-monotonically during the entire sequence. This behavior 
  is not only characteristic of the lowest portion of the spectrum; it is 
  noticed for all values of the index $i$. Simulation data for these two
  materials are shown in Table~\ref{tab:sim}.
  
  \section{Exploiting the sequence evolution}
  \label{sec:evol}
  In the following we present experimental evidence of the
  acceleration of iterative eigensolvers due to the use of approximate
  solutions. We proceed in two stages: first, using a direct
  eigensolver, we compute the eigenpairs of eigenpencil $P^{(\ell)}$
  for a certain fraction of the eigenspectrum. Next, we feed these
  eigenpairs to a selected iterative eigensolver as an initial guess
  when solving for $P^{(\ell+1)}$. Finally we repeat the same
  computation with randomly generated vectors and compare the
  respective CPU times. These tests are repeated for three different
  choices of crystals and at each increasing outer-iteration
  index. The result of the comparison provides a measure of the
  possible speed-up iterative methods can achieve when used in
  sequences of correlated eigenvalue problems.
  
  Among the many available iterative methods, only a few possess the
  ability of taking advantage of multiple approximate solutions and
  gain a sensible speed-up. In this respect an iterative algorithm has
  to comply with two essential properties: 1) the ability to receive
  as input a sizable set of approximate eigenvectors, and 2) the
  capacity to solve simultaneously for a substantial portion of
  eigenpairs.  In particular we considered three classes of iterative
  eigensolvers: Davidson, conjugate gradient and subspace
  iteration. Krylov subspace methods~\cite{Ste,Saad} were not included
  due to their inability to even converge all the requested eigenpairs
  without incurring in ``stalled'' computations. For each class we
  choose a block eigensolver whose characteristics are the closest to
  the requirements listed above.  Block methods generalize the action
  of an eigensolver on one single approximate eigenvector to a set of
  vectors. As such they satisfy quite clearly the first requirement
  while the ability to obtain multiple eigenpairs in one shot is
  rather algorithm dependent.
  
  In this section two sets of numerical results are illustrated: first
  we test three iterative algorithms exclusively in a Matlab
  environment. The objective is to gain an overview of the behavior of
  the eigensolvers and their relative efficiency in exploiting
  approximate solutions.  In a second moment we focus on one of the
  eigensolvers, briefly illustrate its implementation in the C
  programming language and conduct a more performance-oriented
  numerical analysis of its features.
  In section~\ref{sec:block} we illustrate the salient characteristics and properties of the block algorithms under scrutiny. This is followed 
  by a detailed description of the experimental setup. Finally in section~\ref{sec:speed} the results of our numerical tests are
  presented and explained.
  
  
  \subsection{Block iterative eigensolvers}
  \label{sec:block}
  By their own nature iterative eigensolvers come in many variants and
  often some of their internal parameters can be tuned opportunely. In
  block methods one obvious input parameter is the block size \blk;
  since block methods are notoriously more efficient in dealing with
  multiple and clustered eigenvalues, adjusting \blk\ could improve
  the eigensolver performance. On the other hand, knowing a priori the
  positions of clusters of eigenvalues is usually not possible, so a
  value of \blk\ that is optimal for a certain fraction of
  eigenspectrum may not work for another.  On algorithms augmented
  with a polynomial accelerator another evident parameter is the
  polynomial degree \polym.  In this case a good choice for \polym\
  would balance the influence of two competing mechanisms: CPU time
  spent in performing matrix-matrix operations and the efficiency with
  which vectors converge to the solution.
  
  In light of these considerations, it is important to understand that
  an eigensolver that performs well for a certain problem may become
  very slow for another. In particular we want to stress that we
  utilize iterative eigensolvers outside the ``comfort zone'' where
  they usually perform the best, namely sparse eigenproblems. By using
  these solvers on dense eigenproblems we are deliberately pushing
  them to the limit. The idea is to compensate for the misuse of the
  solvers by providing them with approximate eigenvectors. In doing so
  we want to test how effectively each eigensolver can gain in
  performance by exploiting problem solutions of the previous
  outer-iteration.
  
  In order to extract the best behavior, all the numerical tests were
  carried out after optimizing each solver to the specific problem at
  hand. Such an optimization was achieved by tuning the values of the
  available parameters (see table~\ref{tab:pars}). Despite these
  efforts not all eigensolvers seemed to have been able to take full
  advantage of the approximate solutions. We emphasize that this
  behavior is in no way detrimental to the value of the particular
  solver; it just indicates that the chosen eigensolver was too
  stressed to be able to overcome the inherent difficulties of the
  dense eigenproblems selected.  All the eigensolvers described below
  were inherently developed for the lower portion of the
  spectrum. Unless otherwise specified, in the rest of the paper we refer exclusively to this part of the eigenspectrum.\\
  
  \noindent
  \textcolor[rgb]{1,0,0}{\bf Block Chebyshev-Davidson (BChDav) --} This eigensolver is part of the Davidson-like methods. These methods sacrifice the
  Krylov subspace structure by computing an approximate residual which, opportunely preconditioned, is used to augment the subspace.
  In 2007 Saad and Zhou developed a version of the Davidson algorithm for problems where the preconditioning could be
  unknown or too expensive to compute~\cite{Saad-Zhou}; in particular by filtering with Chebyshev polynomials the augmentation 
  vector instead of solving a correction equation, they achieve better numerical results. 
  
  In 2010 Zhou implemented a block version of this method by adding an 
  inner restart loop beside the usual outer restart one~\cite{Zhou}. This version of the method succeeds in better deflating converged vectors 
  and has the added ability to accept approximate solutions in place of the standard augmentation vectors. In particular the inner restart
  loop allows the addition of a new block of approximate vectors as soon as some of the sought after eigenpairs have converged.
  This is the first of the Chebyshev filtered methods that we tested. In his work Zhou has also shown that the method is not particularly
  sensitive to \blk\ nor to \polym. In our test we verified that this is indeed true and fairly small variations in computing time are observed by
  varying one or both of the parameters independently of the size of the eigenproblem. For our test we used a Matlab version of BChDav 
  kindly provided by Y.~Zhou.\\
  
  \noindent
  \textcolor[rgb]{0,0.498,0}{\bf Locally optimized block
    preconditioned conjugate gradient (Lobpcg) --} Developed in 2001
  by Knyazev~\cite{Knya}, this algorithm uses a locally optimized
  version of a three-term recurrence relation for the preconditioned
  conjugate gradient method.  In practice the Rayleigh-Ritz method is
  used for the eigenpencil on a trial subspace generated by the
  current guess for the Ritz vector, the preconditioned residual, and
  a third Ritz vector built by maximizing the Rayleigh quotient.
  Knyazev implemented a block version of the algorithm where the
  three-term relation is generalized for a block of vectors.
  
  We performed our tests using the freely available Matlab version of
  the code. In this version \blk\ is set by construction to be equal
  to the number of requested eigenpairs \nev. While this choice seems
  natural, it encounters difficulties in converging the last few
  vectors requested due to the reduced size of the trial
  subspace. Despite being potentially disruptive, this limit is
  usually overcome by setting a rather large tolerance for the
  residuals and restarting the solver with the obtained solutions and
  a smaller tolerance.  In order to put this method on par with the
  others analyzed, we augmented \blk\ by 5\% of \nev\ and introduced a
  slight modification in the stopping criterion. In this way we could
  use the eigensolver in one run and obtain solutions possessing the
  required accuracy.
  
  We deliberately used this method without a preconditioner, since in
  general such an operator is unknown in DFT-generated sequences
  \seq. Moreover Lobpcg accepts the matrices of a generalized
  eigenproblem directly as input. We show that for the case of
  FLAPW-generated eigenproblems this is a sub-optimal use of the
  solver due to the high conditioning number of the overlap matrices
  $B$.  Much better behavior is observed when the generalized
  eigenproblems are reduced to standard form and then
  inputted to the eigensolver (see Table~\ref{tab:lob}).\\
  
  \noindent
  \textcolor[rgb]{0,0,1}{\bf Chebyshev filtered subspace iteration (ChFSI) --} This is the second of the Chebyshev-filtered methods we tested. 
  This method is well known in the literature~\cite{Saad} and
  has been developed in the context of Density Functional Theory in real-space by Chelikowsky {\sl et al.}~\cite{Tiago-Cheli,TC2} for 
  the PARSEC code~\cite{PARSEC}. Subspace iteration is perhaps one of the oldest known iterative algorithms. 
  This is by nature a block solver since it simply tries to build an invariant eigenspace by block-multiplying a set of vectors 
  with the operator to be diagonalized.
  It is well known that this class of methods converges very slowly and could end up with blocks
  of vectors spanning an invariant eigenspace that are linearly dependent (rank deficient case). 
  By using a polynomial filter on the initial set of \blk\ vectors both these problems are improved very efficiently and the method
  experiences a high rate of acceleration. 
    
  Also in this case the degree of the filter \polym\ does not particularly influence the convergence 
  as long as it is sufficiently high. In general the value for \blk\ is chosen to be bigger than the requested
  number of eigenpairs \nev. This choice ensures that the last eigenpairs converge quickly enough. In ChFSI, the 
  subspace iteration loop is preceded by a Lanczos step in order to determine the boundaries of the interval
  out of which the set of vectors is filtered. 
  
  Starting from the backbone of a routine first developed for a Matlab version of the PARSEC code, we implemented
  a more sophisticated algorithm including an inner loop and a natural ``deflation and locking'' mechanism for the converged eigenvectors.
  Contrary to the code used in PARSEC, our implementation performs several filtering steps until all eigenpairs
  have converged to the required tolerance. A more detailed description of the routine is provided in Algorithm~\ref{alg:ChFSI} below.\\
  
  \begin{algorithm*}[h!t]
  \caption{Chebyshev Filtered Subspace Iteration with locking}
  \label{alg:ChFSI}
  \begin{algorithmic}[1]
  \Require Matrix $H^{(\ell)}$ of the standard problem, approximate eigenvectors $\hat{Y} := \left[ \hat{y}^{(\ell-1)}_1, \ldots, \hat{y}^{(\ell-1)}_{\blk}\right]$ and eigenvalues $\lambda^{(\ell-1)}_1$ and $\lambda^{(\ell-1)}_{\blk + 1}$
  \Ensure Wanted eigenpairs $\left( \Lambda,Y\right).$
  \item[]
  \State Estimating upper bound for the largest eigenvalue \label{lst:line:lanczos} \Comment{{\sc Lanczos}}
  \Repeat
  \State Filtering the vectors $\hat{Y}=C_{\polym}(\hat{Y}).$ by a \polym\ Chebyshev polynomial  \label{lst:line:cheby} \Comment{{\sc Chebyshev filter}}
  \State Re-Orthonormalizing $\hat{Y}.$ \label{lst:line:rrstarts} \Comment{{\sc Rayleigh-Ritz} (Start)}
  \State Compute the Rayleigh quotient $G=\hat{Y}^{\dagger}H^{(\ell)}\hat{Y}$ 
  \State solve the reduced standard problem $Gw=\lambda w$ giving $\big(\hat{\Lambda}, \hat{W}\big).$ 
  \State Compute new $\hat{Y} = \hat{Y}\hat{W}.$ \label{lst:line:rrends} \Comment{{\sc Rayleigh-Ritz} (End)}
  \For{$i=\textrm{converged} \to \textsc{nev}$ } \label{lst:line:resstarts} \Comment{{\sc Deflation \& Locking} (Start)}
    \If{$\cfrac{||H^{(\ell)}\hat{Y}(:,i)-\hat{\Lambda}(i)\hat{Y}(:,i)||}{||\hat{Y}(:,i)||}<\textsc{tol}$} \hfill {\small Checking residuals of Ritz vectors.}
    \State $\Lambda\gets \left[\Lambda\;\hat{\Lambda}(i)\right]$ \hfill {\small Locking converged eigenpairs.}
    \State $Y\gets \left[Y\;\hat{Y}(:,i)\right]$\label{lst:line:lock-end} \label{lst:line:resends} 
    \Else
    \State \textbf{break}
    \EndIf
    \EndFor \Comment{{\sc Deflation \& Locking} (End)}
    \State converged $\gets i$
    \State $\hat{\Lambda} \gets \left[\hat{\Lambda}(\textsc{converged}:\textsc{end})\right]$
    \State $\hat{Y}\gets \left[\hat{Y}(:,\textsc{converged}:\textsc{end})\right]$
  \Until{ converged $<$ \textsc{nev}}
\end{algorithmic}
\end{algorithm*}


  \subsection{Experimental setup}
  \label{sec:setup} 
  In order to test the behavior of the selected solvers, we singled out sequences of eigenproblems arising from
  three DFT systems presenting heterogeneous physical properties: a metal, an ionic crystal and a high-temperature
  superconductor. Besides having different physical properties, these systems differ in the size of the eigenpencils, 
  and in the structure of the eigenspectrum. Since the efficiency of 
  iterative eigensolvers is often quite sensitive to these characteristics, we could verify the robustness of our
  conclusions in different conditions. Simulation data are collected in Table~\ref{tab:sim}.
  
  Each sequence \seq\ of eigenproblems was generated by running
  simulations using FLEUR, a FLAPW-based code developed in
  J\"ulich~\cite{FLEUR}.  By fixing a specific \kv-vector we
  identified one sequence per system and stored the relative
  $A^{(\ell)}$ and $B^{(\ell)}$ matrices.  For all the simulations we
  adopted the value $\eta < $1e-03 (see eq.~(\ref{eq:cycle})) as a
  signal for convergence; in turn this choice determines the maximum
  value of the iteration index for each sequence. All simulations were
  run on JUROPA, a powerful cluster-based computer operating in the
  Supercomputing Centre of the Forschungszentrum J\"ulich.
  
  The first set of numerical tests was performed using Matlab, version
  R2011b (7.13.0.564) under an OpenSuSE 12.1, running on two Intel
  i7-870 (Nehalem ``Lynnfield") quad-core processors at 2.93 GHz. In
  order to avoid lengthy simulations, we took advantage of Matlab
  multi-threaded routines so that up to four cores and 8 Gb of RAM (2
  DDR3, 1333 MHz) were fully dedicated to computations.  The second
  set of tests for the C language implementation of ChFSI was
  performed on one node of JUROPA equipped with 2 Intel Xeon X5570
  (Nehalem-EP) quad-core processors at 2.93 GHz and 24 GB memory
  (DDR3, 1066 MHz).  All CPU times were measured by running each test
  for each algorithm 12 times and taking the median of the results.
  
  Directly solving for the generalized eigenproblems makes a fair
  comparison among the selected eigensolvers cumbersome.  One of the
  solvers can use $B$ directly (i.e.~Lobpcg) while the other two deal
  with generalized eigenproblems by left multiplying $B^{-1}$ to $A$.
  Unfortunately, because of the over-completeness of the set of basis
  functions in eq.~(\ref{eq:expon}), the overlap matrices $B$ are
  positive definite but usually quite ill-conditioned. This
  characteristic prevents the inversion of $B$ due to inherent
  numerical difficulties associated with a few singular values of $B$
  very close to zero.  Consequently, in order to set the solvers on
  equal footing, we prepared our tests reducing all the eigenpencils
  to standard eigenvalue problems.  By using the Cholesky
  decomposition of $B=LL^T$, we defined $H = L^{-1} A L^{-T}$ and
  solved for $\tilde{P}: H y = \lambda y$, with $y=L^Tx$.  This choice
  solves the problem of computing $B^{-1} A$ for both BChDav and
  ChFSI, but does not justify a priori the use of $H$ for Lobpcg.
  
  In order to address this disparity in the use of Lobpcg we
  investigated its performance when solving directly for the
  generalized eigenproblem versus the combined time spent in reducing
  the problem to a standard one before feeding it to this solver. The
  results are shown in Table~\ref{tab:lob}. In the column labeled
  ``GEN Problem'' the CPU time to completion is listed for the eigensolver
  when the two matrices $A$ and $B$ are directly inputted. In the
  column ``STD Problem" the CPU time is the sum of the reduction of
  the generalized eigenproblem to standard form $H$, plus the solution
  of the standard eigenproblem by the solver. The tests were executed
  on the CaFe$_2$As$_2$ system (n=2612) for iter=17, \nev=136,
  \blk=142, {\tt maxiter}=2000\footnote{See section~\ref{sec:speed}
    for a detailed explanation of the parameters meaning.}.  All the
  numbers refer to median values over 10 repetitions performed for
  both random vectors and approximate solutions. The results clearly
  show that as the solution accuracy grows, Lobpcg increasingly
  struggles to solve the generalized eigenproblems directly: a direct
  consequence of the ill-conditioned nature of $B$.  Consequently, as
  for the other algorithms, we performed numerical tests for Lobpcg on
  just the reduced problem $\tilde{P}: H y = \lambda y$.
  
  \begin{table}[ht]
  \caption{Lobpcg -- Generalized vs. Standard eigenproblems}
  \vspace{0.2cm}
  \centering
  \begin{tabular}{c c r r }
  \hline \hline \\[-3mm]%
  Accuracy & \shortstack{Initial\\ vectors} & \shortstack{CPU time (sec)\\ GEN Problem} & \shortstack{CPU time (sec)\\ STD Problem} \\ [0.5ex]
  \hline \\[-3mm]%
  \multirow{2}{*}{{\tt TOL}:1e-06} & random & 763.05 & 32.73\\ %
  	& approx. & 13.07 & 11.55\\[0.5ex]
	 \hline\\[-3mm]%
  \multirow{2}{*}{{\tt TOL}:1e-07} & random & 918.47 & 42.97\\ 
   	& approx. & 48.66 & 15.38\\[0.5ex]
	 \hline \\[-3mm]%
  \multirow{2}{*}{{\tt TOL}:1e-08} & random & 1287.60 & 48.90\\ 
  	& approx. & 149.22 & 20.46\\
  \end{tabular}\\
  \label{tab:lob}
  \end{table}
      
  For each sequence of eigenproblems we tested the eigensolvers for all outer-iteration indices 
  and the choices of \nev\ required by the DFT simulation (see Table~\ref{tab:sim}). For 
  each of these choices the numerical experiments can be schematically divided into four stages:
  \begin{enumerate}
	\item solving the standard problem $H^{(\ell)} y = \lambda\ y$ using a direct method (i.e.~the MRRR algorithm) 
  	and storing a fraction of the eigenvectors in a matrix $Y^{(\ell)}$;
	\item solving $H^{(\ell+1)} y = \lambda\ y$ utilizing the iterative eigensolvers with randomly generated vectors and 
	recording the CPU time to completion, t$_{\rm rnd}$;
	\item solving $H^{(\ell+1)} y = \lambda\ y$ utilizing iterative eigensolvers with the eigenvectors in $Y^{(\ell)}$ and 
	recording the CPU time to completion, t$_{\rm app}$; 
	\item comparing the CPU times measured at 3. and 4. by plotting the speed-up, $\frac{\rm t_{\rm rnd}}{\rm t_{\rm app}}$. 
  \end{enumerate}
  
  In the Matlab tests, speed-up and t$_{\rm app}$ are plotted
  separately with respect to the iteration index for all three
  eigensolvers. On the one hand we expect to observe a growth in
  speed-up as the outer-iteration index increases; eigenvectors become
  more and more collinear as the sequence progresses (see
  Figure~\ref{fig:evol}), a behavior that should enhance the
  efficiency of the iterative eigensolvers.  On the other hand, from
  the absolute time t$_{\rm app}$ plots, we gain some perspective on
  which algorithm, among the three, may have a performance advantage.
  To conclude we focus on the ChFSI algorithm, program a C language
  version, and test its speed-up, performance and scalability.
  
  
  \subsection{Numerical results and discussion}
  \label{sec:speed}
  Once more we would like to stress that we tested a range of values
  for all the parameters available to each eigensolver.  The aim was
  to optimize the algorithms for the sequences of problems at hand
  when started with random vectors before endeavoring in the task of
  evaluating the eigensolvers' speed-up.
  Once determined, the optimal set of parameters was maintained for all the sequences of the physical systems under scrutiny. 
  A schematic list of these values is available in Table~\ref{tab:pars}. In this table
  {\tt nkeep} is a parameter indicating the number of vectors that are kept after a restart of the inner loop. {\tt vimax} and 
  {\tt vomax} are the maximum size of the augmentation subspace for the inner and outer loop (not to be confused with the outer-iteration DFT cycle)
   of BChDav, while the {\tt maxiter} sets
  the maximum number of iterations for the inner loop. {\tt lanczos-iter} indicates the maximum number of iterations of 
  the Lanczos step that ChFSI uses to bound the spectrum to be filtered out; this number can vary substantially depending on the
  nature (random or approx.) of the starting vectors. All tests for each eigensolver were run requiring the same accuracy for the 
  residuals~\footnote{While BChDav uses relative residuals, for both Lobpcg and ChFSI absolute residuals are employed. Since $||H||$ 
  was in most cases of the order of 10, differences between the use of two definitions of residuals are not very significant.}. 
  
  \begin{table}[ht]
  \caption{Parameter settings}
  \vspace{0.2cm}
  \centering
  \scalebox{0.83}{
  \begin{tabular}{| c || c c c |}
  \hline \hline 
  \shortstack{\, \\Parameters} & {\bf BChDav} & {\bf Lobpcg} & {\bf ChSI}\\ [0.5ex]
  \hline 
  \blk\ & 35 & {\small 1.05$\times$\nev} & {\small \nev$+$40} \\ \hline%
  \polym\ & 25 & -- & 25 \\ \hline%
  {\tt nkeep} & 3$\times$\blk & \blk$-${\tt nconv} & \blk$-${\tt nconv} \\ \hline%
  {\tt vimax} & {\small max($\lceil\frac{\nev}{4}\rceil$, 5$\times$\blk, 30)} & -- & -- \\ \hline%
  {\tt vomax} & {\small \nev $+$50} & -- & -- \\ \hline%
  {\tt maxiter} & {\small max($\lfloor\frac{n}{2}\rfloor$, 300)} & 1000 & -- \\ \hline%
  {\tt lanczos-iter} & -- & -- & {\small 10 $\div$ 3$\times$\blk} \\ \hline%
  {\tt TOL} & $10^{-10}$ & $10^{-10}$ & $10^{-10}$\\ \hline%
  \end{tabular}
  }
  \label{tab:pars}
  \end{table}

  
\subsubsection{Matlab tests}
\label{sec:matlab}
\begin{figure}[!htb]
  \centering
	\subfigure[Speed-up]
	{\hspace{-0.4cm}\includegraphics[scale=0.22]{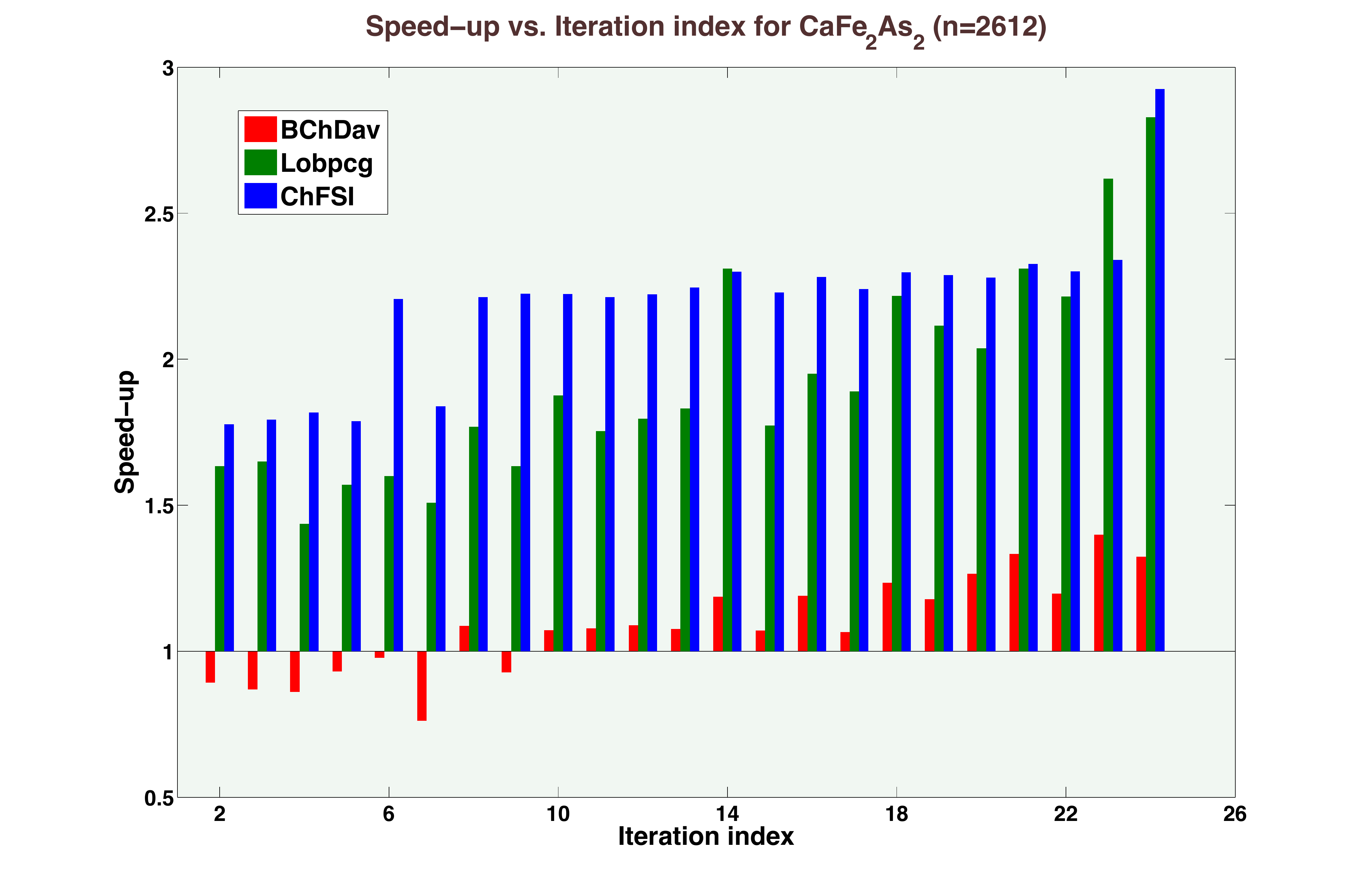}}
	\subfigure[Time to completion]
	{\hspace{-0.4cm}\includegraphics[scale=0.23]{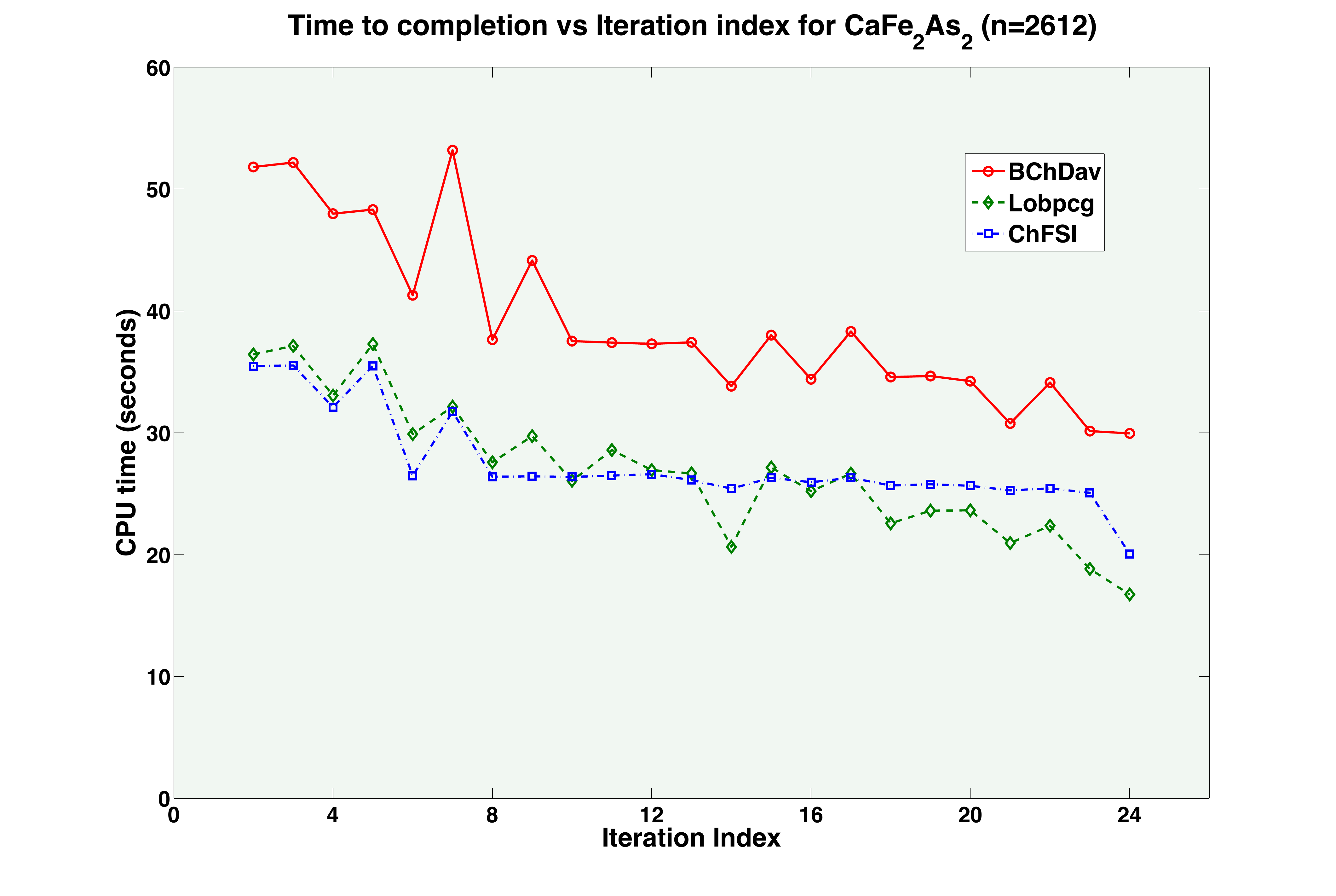}}
	\caption{\em Comparison between the 3 most effective block iterative eigensolvers for CaFe$_2$As$_2$ with respect 
	to the outer-iteration index.}
	\label{fig:CaFe}
  \end{figure}

\begin{figure}[!htb] 
  \centering
	\subfigure[Speed-up]
	{\hspace{-0.3cm}\includegraphics[scale=0.22]{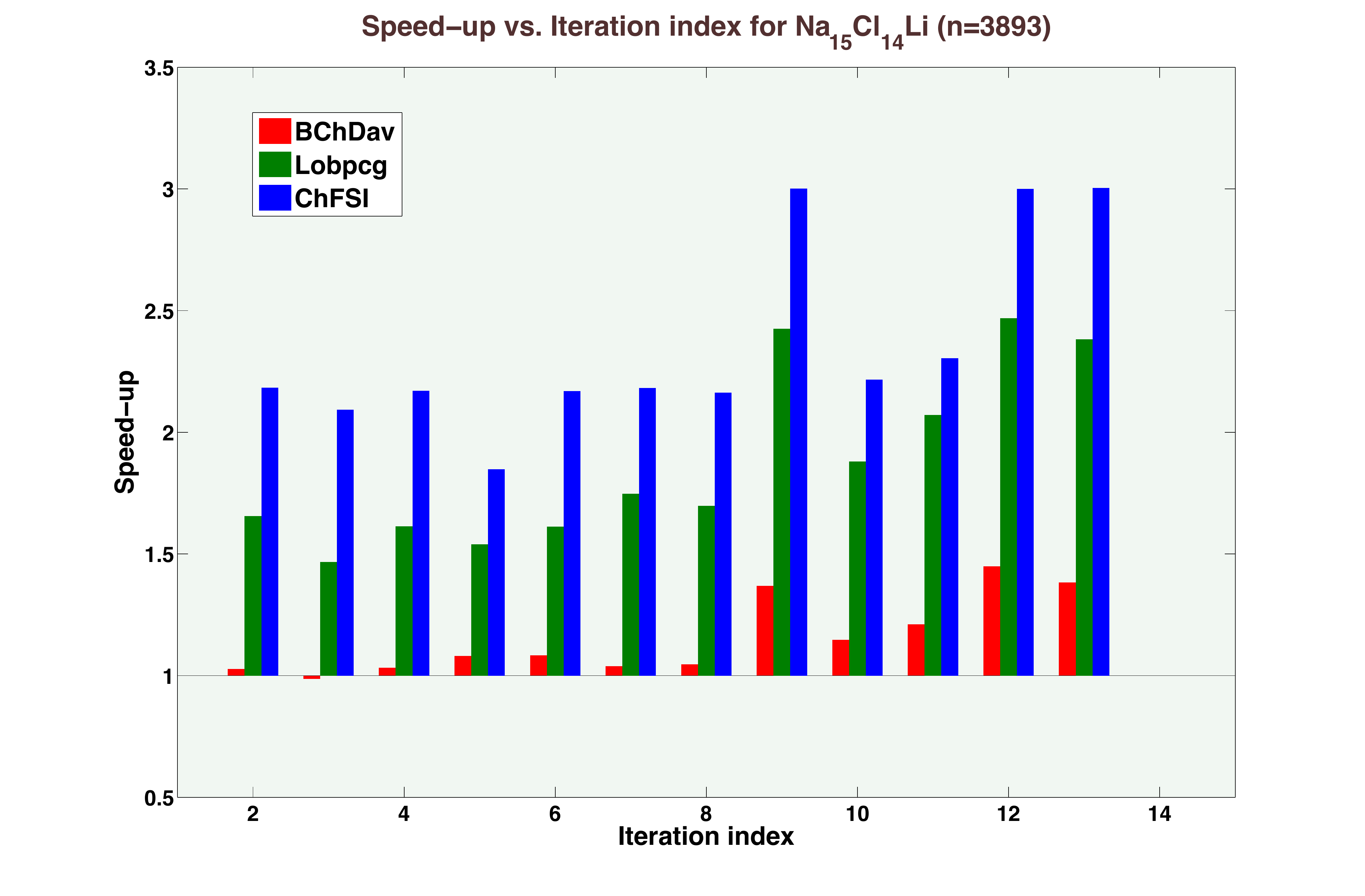}}
	\subfigure[Time to completion]
	{\hspace{-0.3cm}\includegraphics[scale=0.23]{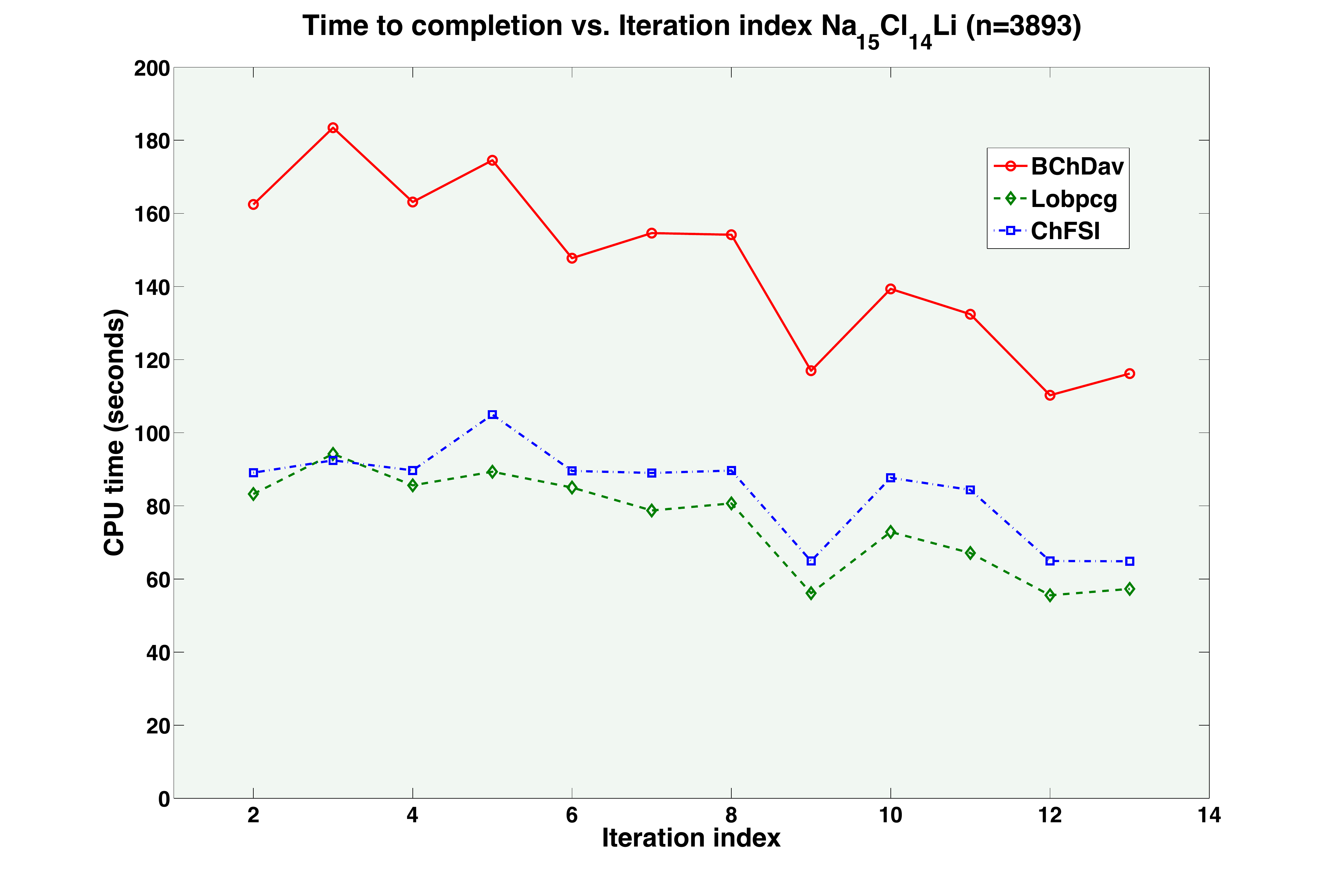}}
	\caption{\em Comparison between the 3 most effective block iterative eigensolvers for Na$_{15}$Cl$_{14}$Li with respect to 
	the outer-iteration index.}
	\label{fig:NaCl}
  \end{figure}

In Figures~\ref{fig:CaFe}\ and \ref{fig:NaCl} we have plotted the
speed-up and CPU times for all three eigensolvers BChDav, Lobpcg and
ChFSI.  We aim at visualizing the improvement in performance of each
algorithm and, at the same time, gaining some insights on the behavior
of the different methods. Due to the limited memory available on the
desktop machine running Matlab, we measure the execution time only for the
two smallest physical systems listed in Table~\ref{tab:sim}.  In
Figure~\ref{fig:CaFe} the data refers to the lowest 136 eigenpairs of
the CaFe$_2$As$_2$ system, corresponding to 5.2\% of the eigenvalue
spectrum. In Figure~\ref{fig:NaCl}\ the plots pertain to the lowest
256 eigenpairs of the $n=3898$ Na$_{15}$Cl$_{14}$Li system,
corresponding to the lowest 6.6\% of the spectrum. In order to clearly
address the performance of each algorithm, the numerical results on
the plots are discussed independently for each solver.
  
  \noindent
  \textcolor[rgb]{1, 0, 0}{\bf BChDav} -- In both figures the behavior
  of this algorithm is clearly influenced by the outer-iteration
  index; in the first half of the sequence the solver does not gain
  almost any advantage from the use of approximate solutions and only
  in the second half it does experience a limited speed-up, never
  reaching values higher than 1.5X. In particular, plot (a) of
  Figure~\ref{fig:CaFe} shows the solver is actually penalized by the
  approximate eigenvectors and slows down at the beginning of the
  sequence. This anomalous behavior of BChDav could largely be caused
  by its non-optimal use.  In particular the Chebyshev filtering step
  could be aligning more than one approximate eigenvector along the
  same direction leading to a rank deficient subspace. This fact in
  combination with the modest size of the matrices may explain the
  negative speed-up for the first outer-iterations.
  
  For a larger matrix system (see 
  plot (a) of Figure~\ref{fig:NaCl}) we again observe that BChDav does not take much advantage of the approximate 
  solutions at the beginning of the sequence but it is at least not penalized by them. 
  Despite its limited speed-up, CPU timings decrease substantially as the iteration index grows when the solver is fed approximate vectors.
  In turn this observation indicates that the correlation among eigenvectors in the sequence has nonetheless a certain influence on the
  behavior of the eigensolver.     
   
  \noindent
  \textcolor[rgb]{0, 0.498, 0}{\bf Lobpcg} --
  Quite different is the performance of Lobpcg: this solver has speed-ups 
  larger than 1.5X already at the beginning of the sequence. 
  Both figures show that as the sequence progresses, Lobpcg speed-up presents moderate oscillations
  instead of a steady increase. 
  Moreover while it reaches more than 2.5X for the last two outer-iteration indices of the CaFe$_2$As$_2$
  system, it never does so for the larger system. The oscillation could be the consequence of different 
  eigenvalue clustering for eigenproblems of distinct iteration indices.
  The decrease in speed-up for the larger system can be
  explained by the effectiveness of the conjugate gradient method to converge vectors
  even when Lobpcg is inputted random vectors. Conversely this trend implies
  a decreasing threshold for the speed-up for larger matrix sizes as signaled by
  the flattening of the CPU time curve in going from Figure~\ref{fig:CaFe} to Figure~\ref{fig:NaCl}.
  This last observation suggests that, despite the nice speed-up, Lobpcg may take less and less
  advantage of approximate solutions as the size of the eigenproblem increases. 
  
  \noindent
  \textcolor[rgb]{0, 0, 1}{\bf ChFSI} -- Among all the eigensolvers
  ChFSI seems to be able to take greater advantage of the approximate
  solutions and have a more predictable behavior: in plot (a) of
  Figure~\ref{fig:CaFe} its speed-up proceeds in 3 different steps
  jumping from $\sim1.8$X to $\sim2.2$X and to almost 3X at the end of
  the sequence. In Figure~\ref{fig:NaCl} the same trend appears in
  just two steps: ChFSI speed-up starts at $\sim2$X and reaches just
  above 3X towards the end of the sequence. In the next subsection we
  explain in more detail the reasons for this step-like behavior.  For
  the moment it is enough to say that ChFSI exploitation of
  approximate solutions seems to increase with the matrix size. This
  aspect together with the fact that its CPU time curve is not very
  different from Lobpcg's delivers a better promise for this
  eigensolver to be the optimal one to take advantage of the
  correlation magnification as the sequence progresses. With this last
  consideration in mind we decided to single out ChFSI and implement
  it in the more efficient C programming language in order to be able
  to evaluate its performance and scalability.
  
  We conclude this small section reiterating that, despite the differences in their behavior, all eigensolvers
  considered succeed to exploit the eigenvectors increasing collinearity, confirming the expectations that
  motivated this study.
  
  
  \subsubsection{The performance of ChFSI}
  \label{sec:ChFSI}
  The numerical tests of this section have their origin in an C
  language implementation of ChFSI that, for sake of clarity, we will
  refer to as ChFSIc.  Besides being able to carry out performance
  tests, this implementation also allows us to run ChFSIc on JUROPA, a
  J\"ulich-based general purpose cluster with a large memory per node
  (24 Gb). We are able now to run the same tests of
  Section~\ref{sec:matlab} on the larger matrix size systems of
  Table~\ref{tab:sim}. The immediate objective is to verify the claim
  we made over the proportionality between eigenproblem size and
  effectivity of the algorithm to exploit approximate solutions.
  
\begin{figure}[!htb]
  \centering
	\subfigure[Speed-up for Na$_{15}$Cl$_{14}$Li.]
	{\hspace{-0.6cm}\includegraphics[scale=0.23]{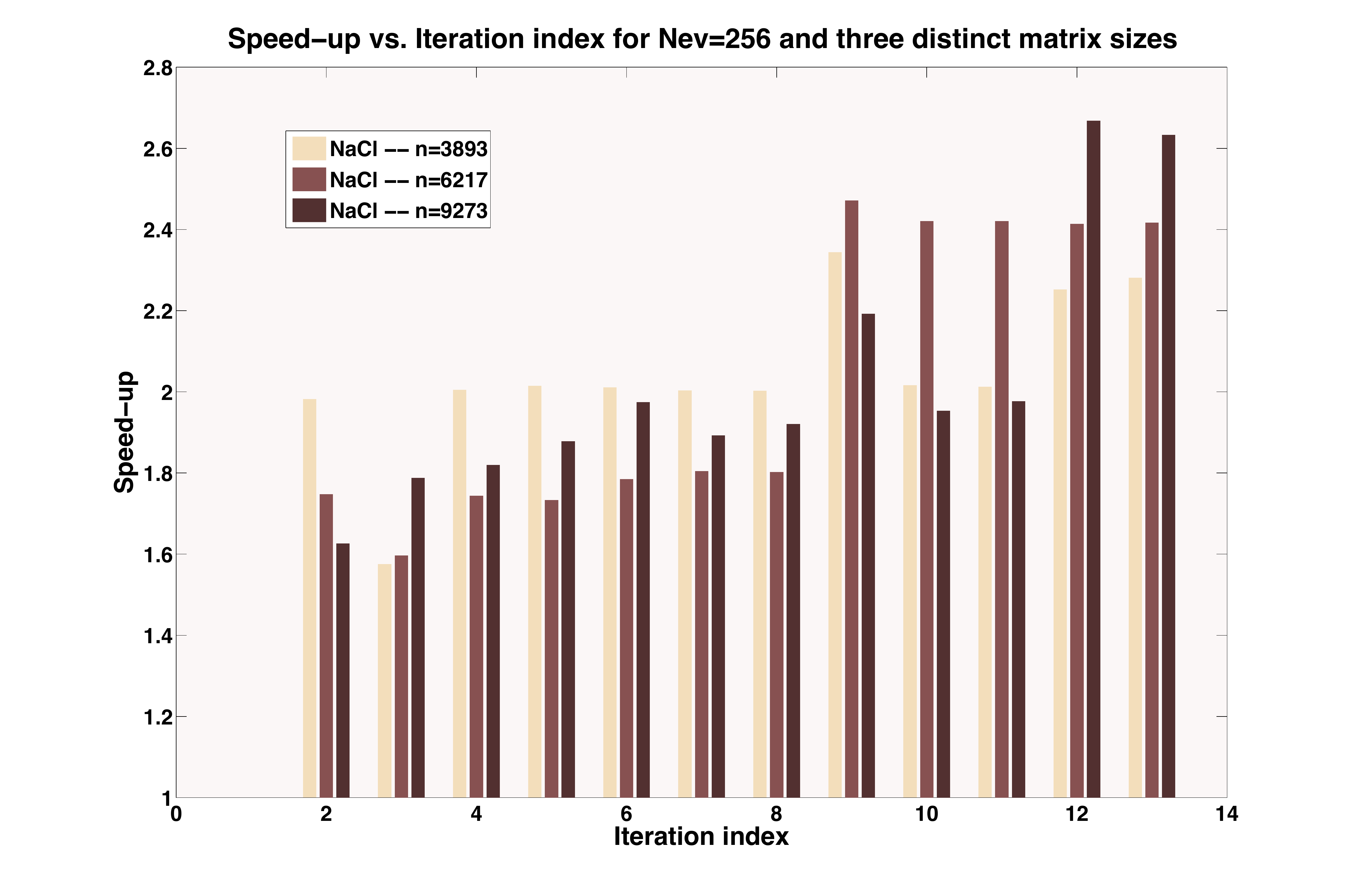}}
	\subfigure[Speed-up for Au$_{98}$Ag$_{10}$.]
	{\hspace{-0.6cm}\includegraphics[scale=0.23]{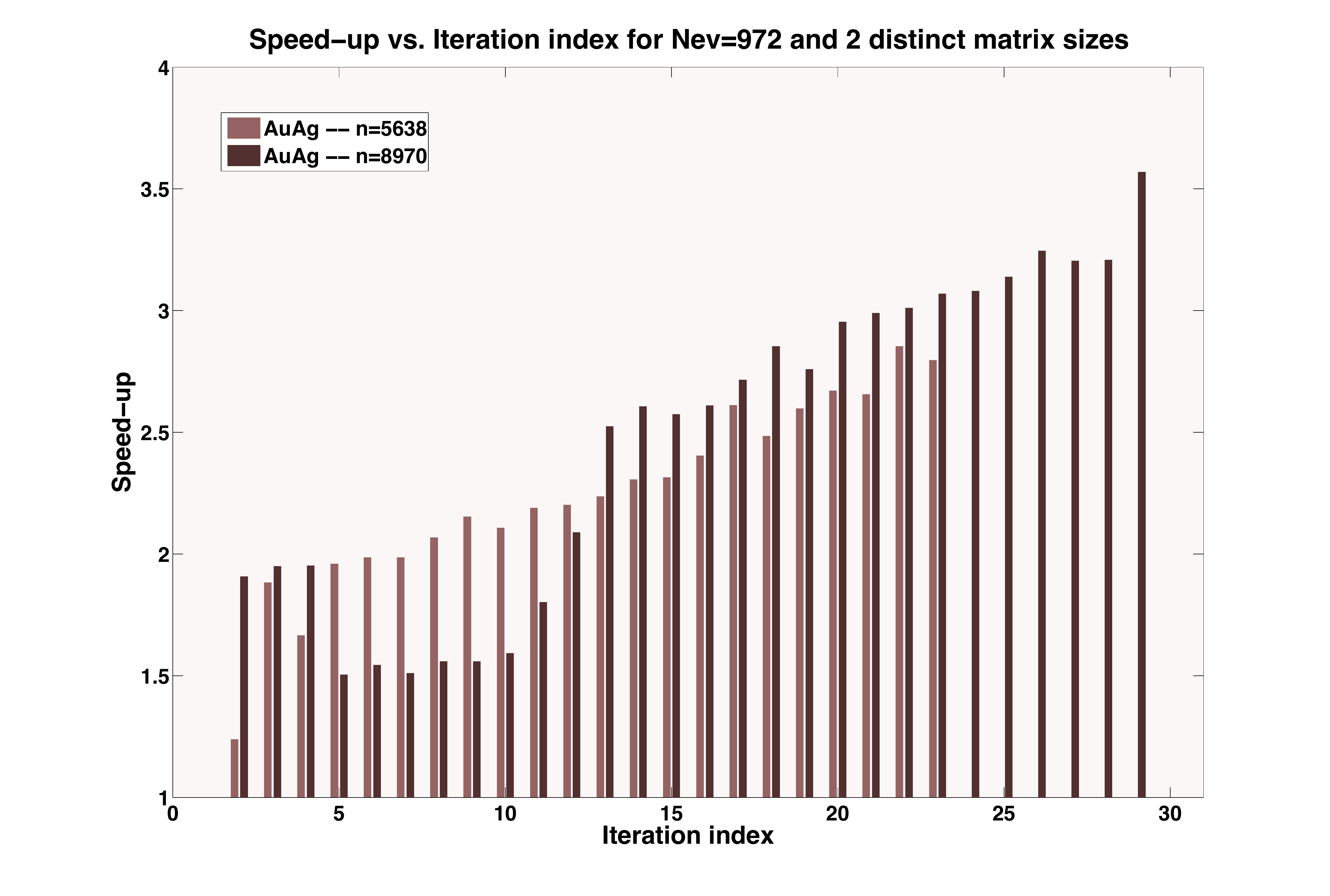}}
	\caption{\em Speed-up for distinct matrix sizes. The matrix sizes are determined by distinct choices for {\bf K}$_{\rm max}$: namely 
	{\bf K}$_{\rm max}$ = $[3.0\ 3.5\ 4.0]$ for Na$_{15}$Cl$_{14}$Li, and {\bf K}$_{\rm max}$ = $[3.0\ 3.5]$ for Au$_{98}$Ag$_{10}$.
	The speed-up is plotted with respect to the outer-iteration index.}
	\label{fig:C_speedup}
  \end{figure}

  Figure~\ref{fig:C_speedup} clearly shows a tendency for the speed-up to grow as the size of the matrices increases. A closer analysis 
  suggests that the extent of the speed-up is not only influenced by the iteration index, but additionally by the sought-after eigenspectrum fraction.
  From both plots it seems that both factors play a not-easy-to-predict role in the trend towards higher effectivity. A closer look at the algorithm 
  (see Algorithm~\ref{alg:ChFSI})
  reveals that speed-up is strongly related to the number of inner-loops that are necessary for all the eigenpairs to converge. Experimentally we observed
  that the total number of loops required by any eigenproblem is well below 10 and tends to decrease suddenly by one or more units every
  few outer-iteration cycles. For larger eigenproblem sizes this transition typically happens at later outer-iteration indices. 
  For instance in plot (a) of Figure ~\ref{fig:C_speedup} the number of inner loops goes from 4 to 3 at the 
  ninth iteration for $n=6217$, while the same change happens at the twelfth 
  iteration for $n=9273$.
  
%
  \begin{table}[ht]
  \caption{Number of inner loop and filtered vectors w.r.t. iteration index for $n=8970$}
  \vspace{0.2cm}
  \centering
  \scalebox{0.7}{
  \begin{tabular}{| c || c | c | c | c | c | c | c | c | c |}
  \hline \hline 
  \shortstack{Iteration \\Index} & 11 & 12 & 13 & $\dots$ & 17 & 18 & 19 & 20 & 21 \\
  \hline 
  \shortstack{Number \\ of loops} & 6 & 6 & 5 & $\dots$ & 5 & 4 & 5 & 4 & 4 \\ \hline%
  \shortstack{Number of\\filtered vectors} & 2794 & 2366 & 1999 & $\dots$ & 1839 & 1766 & 1807 & 1711 & 1680 \\ \hline%
  \end{tabular}
  \label{tab:ladder}
  }
  \end{table}
  In the same figure plot (b) follows the same trend of plot (a) with
  the difference that the speed-up amplification is more gradual than
  step-like.  Consider for example the $n=8970$ case, and observe the
  increase in speed-up from iteration index 10 on with the help of the
  data printed in Table~\ref{tab:ladder}.  We can notice that there
  are two jumps in the number of loops: a sharp one between indices 12
  and 13 where the total number of filtered vectors decreases suddenly
  by a factor of more than 200, and a less definite one across the
  range of indices 17-20. In the latter the total number of filtered
  vectors does not change so dramatically. In practice progressively
  less vectors are filtered during the fifth loop and progressively
  more by the fourth loop. There is not a sharp transition here
  probably due to the absence of a large block of vectors that
  suddenly succeed to converge during the fourth loop.  This behavior
  of the algorithm does not make it less effective but indicates that
  eigenproblems appearing in materials without an energy gap can
  affect the behavior of the filter.
  
\begin{figure}[!htb]
  \centering
	\subfigure[CPU time for sequential fractions of the algorithm. The fractions refer to computation on the n=8970 system.]
	{\hspace{-0.8cm}\includegraphics[scale=0.25]{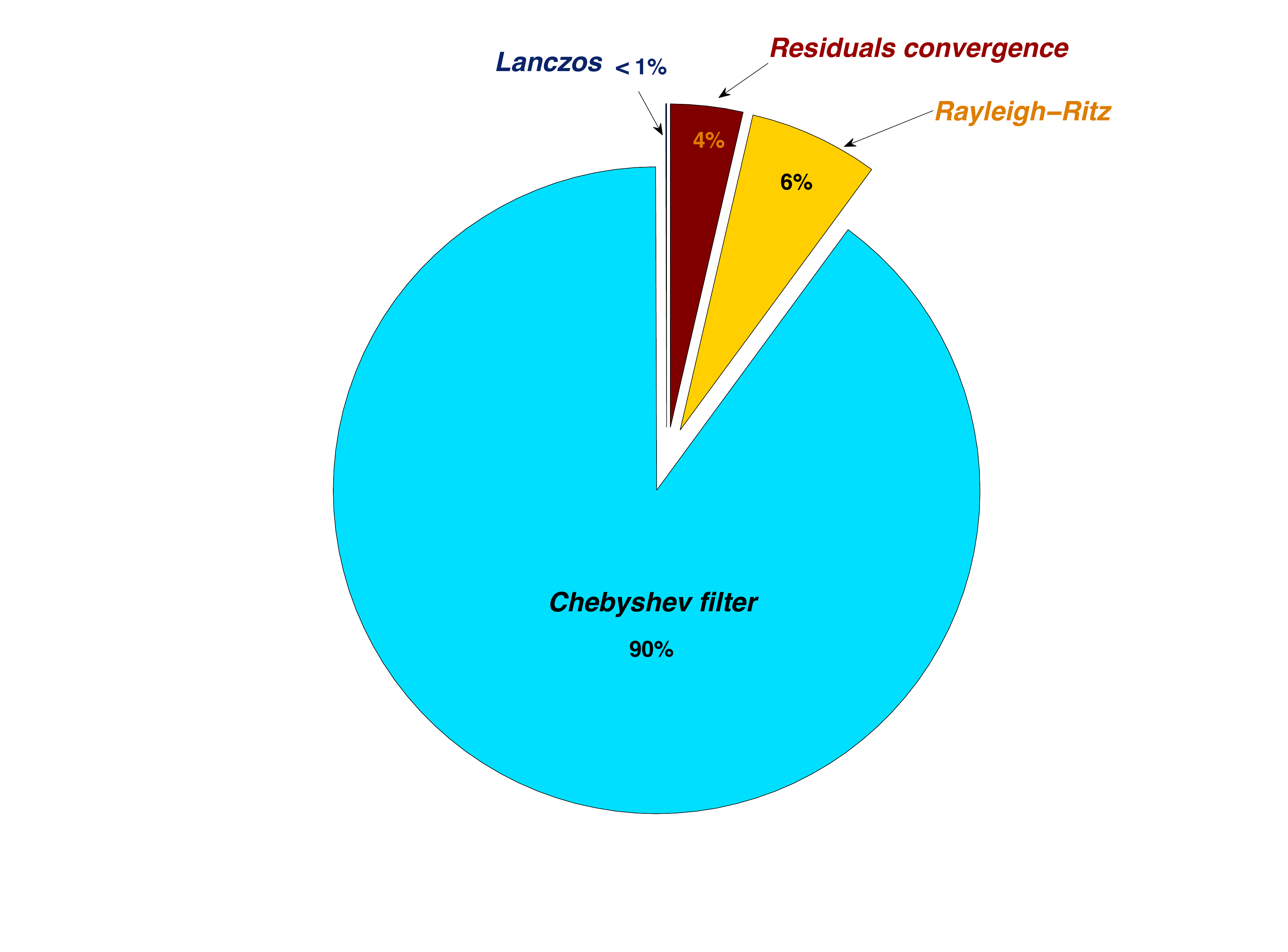}}
	\subfigure[Strong scalability on one computing node using multi-threaded BLAS.]
	{\hspace{-0.8cm}\includegraphics[scale=0.24]{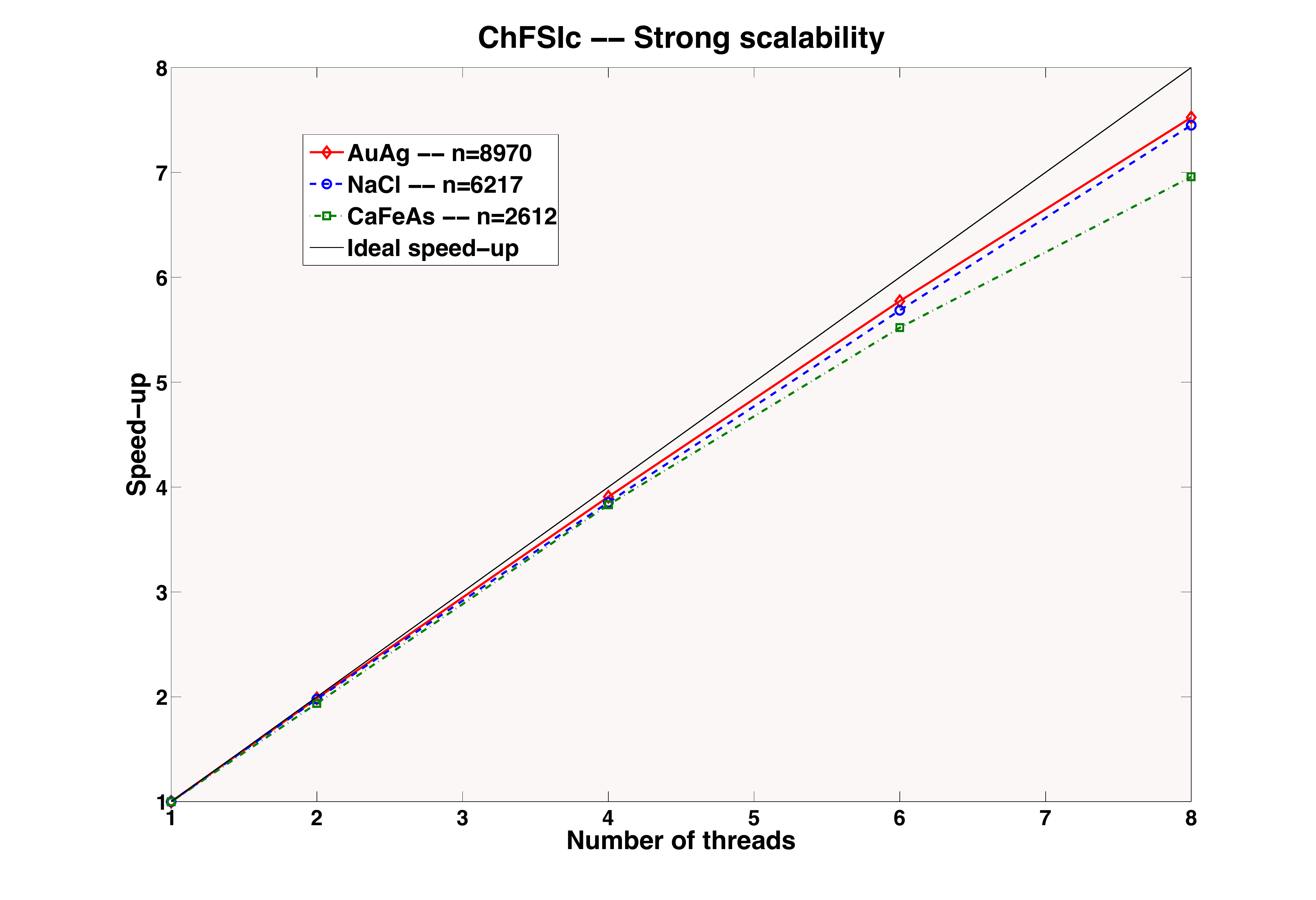}}
	\caption{\em The ChFSIc promise for optimal scalability. All measurements refer to the use of the algorithm when inputted approximate solutions.}
	\label{fig:C_scala}
  \end{figure}

 Since realistic utilization of an iterative eigensolver in a FLAPW-based code requires evidence of scalability when used on multiple cores, 
 we proceeded to test ChFSIc in a final set of numerical tests. 
 From the pie chart in Figure~\ref{fig:C_scala} one can observe that the great majority of computational time for the sequential code is
 spent on the Chebyshev filter fraction (see Algorithm~\ref{alg:ChFSI}). The data refers to the algorithm as used with approximate solutions and the
 fractions of computational time remain pretty much the same for any system tested. In turn this chart suggests that parallelizing the filter is the
 first priority in order to scale on many cores. Since the Chebyshev filter uses almost entirely BLAS-3 kernels, it seems natural to call a 
 multi-threaded version of BLAS.
 
 For our tests we used MKL BLAS version 11.0 (Intel compiler version 12.0.3) over the possible range of cores available on one node of JUROPA.
 In plot (b) of Figure~\ref{fig:C_scala} speed-up of total execution of ChFSIc w.r.t. the number of cores is illustrated for  three larger increasingly eigenproblems. As expected, the larger the
 physical systems the better it scales over an increasing number of cores. What is remarkable is the efficiency of the algorithm: due to the repetitive
 use of the ZGEMM routine by the polynomial filter, its scalability is as close to ideal as one can hope for. 
 Clearly we do not claim this is the whole story, but it is suggestive of
 what ChFSIc could deliver when opportunely parallelized for a larger number of cores. Ultimately such a parallelization would allow the community
 of computational materials scientists to simulate larger and larger systems: an objective that is difficult to achieve with the present implementations
 of FLAPW-based codes.
  
  \section{Summary and conclusions}
  \label{sec:sum}
  Sequences of generalized eigenvalue problems emerge in many common applications. In DFT they arise quite naturally in the search
  for a self-consistent solution for a set of coupled non-linear eigenvalue equations. In the FLAPW method, the matrix pencils making up
  each sequence are dense and are generally solved in isolation using direct solvers straight out of standard libraries. In this paper we started off
  from the idea that a sequence of eigenproblems should be considered as a whole and solved accordingly. The final aim is to uncover
  the possibility of a different approach where the connecting factor between eigenproblems can be exploited to accelerate the solution
  of the entire sequence.
  
  To this end we illustrate the existence of the correlation between
  eigenpencils: subspace angles between eigenvectors of adjacent
  problems decrease monotonically as the sequence progresses towards
  convergence of the FLAPW DFT self-consistent cycle. In other words,
  eigenproblems of growing outer-iteration indices enjoy increasingly
  collinear eigenvectors. Our approach focuses on exploiting this
  property of the sequence by employing an iterative eigensolver
  tailored to accept the solution of an eigenproblem at a certain
  iteration to solve the eigenproblem at the next one. Out of the
  several variants of currently available iterative methods we
  selected three promising eigensolvers and proceeded to devise
  numerical experiments to test the alternative approach. In the
  context of the FLAPW method this is a novel and unexplored
  methodology.
  
  The block solvers we evaluate represent three different classes of
  iterative methods: Davidson, conjugate gradient and subspace
  iteration methods. We carried out an exhaustive series of CPU time
  measurements for each algorithm. Times to reach solution were taken
  initializing the solvers with both random vectors and approximate
  solutions with respect to the outer-iteration index.  Each
  measurement was performed for eigenproblems of different sizes, a
  5-10\% range of sought-after eigenspectrum, and sequences \seq\ of
  increasing length.  Speed-ups were plotted as the ratio between
  ``random'' over ``approximate'' CPU times.
  
  The numerical results show that all eigensolvers take an increasing
  advantage of the approximate solutions as the outer-iteration index
  increases.  While BChDav is probably over penalized by its use on
  dense eigenproblems, Lobpcg and ChFSI seem to be able to
  counter-balance their non-optimal use by substantially increasing
  their performance. In particular both solvers experience a speed-up
  larger than 2.5X towards the end of the sequence. These results
  carry evidence that a different approach to solve sequences of
  correlated eigenproblems arising in DFT is not only possible but
  desirable.
  
  Among the three solvers ChFSI seems to have an extra performance edge thanks to a consistent 
  increase of effectivity for larger sized systems. This result is confirmed by numerical tests run on a C language 
  version of the algorithm and it is further 
  strengthened by the promise for excellent scalability of this eigensolver: when the solver is run in a multi-threaded BLAS environment
  it scales increasingly better for larger sized eigenproblems reaching an efficiency above 85\%. Since the size of the eigenproblems
  depends linearly on the number of atoms, optimal scalability is the crucial element which enables 
  the use of a larger number of processors in order to maximally expand the size of the physical system that can be simulated~\cite{Snyder}.  
    
  In conclusion we observe that out of three iterative block eigensolvers, two of them greatly benefit from the use of approximated solutions.
  This result indicates the possibility of a different strategy in solving sequences of dense eigenproblems  in the context of DFT. Instead of
  using a direct solver for each single eigenproblem in isolation it could be more efficient to exploit the numerical properties of the
  sequence as a whole. Eventually this change in strategy could lead to substantial speed-up of the entire self-consistent 
  outer-iterative process.
  
  While our conclusions constitute a proof of concept, they do not claim to be an exhaustive performance analysis. Some of the described
  algorithms are already available on working platforms more appropriate for performance studies~\cite{PRIMME}. Moreover, current
  efforts are underway to parallelize ChFSIc for shared, distributed and hybrid architectures and a thorough study on its performance will
  be presented in a follow-up publication.
  Eventually, our goal is to conclusively demonstrate that, despite the dense nature of DFT eigenproblems, a promising block
  iterative eigensolver like ChFSIc
  together with an appropriate strategy to reuse previous solutions, can be competitive with direct solvers when only a fraction of the
  eigenspectrum is sought after. 
  
  The authors would like to thank Prof.~Bientinesi for his support, collaboration and steady flow of suggestions, Dr.~Wortmann for his 
  prompt support with the use of FLEUR and the provision of the physical systems for the numerical tests, Prof.~Zhou for kindly providing
  one of the Matlab routines used in running the numerical tests, Prof.~Stathopoulos for his suggestions and finally 
  Prof.~Bl\"ugel for his enthusiasm and support for our ideas. A special acknowledgement to the J\"ulich Supercomputing Centre
  and the John von Neumann Institute for Computing for providing the computational resources to perform all numerical experiments.
  
  \bibliographystyle{plain}

\end{document}